\documentclass[paper=A4,11pt,DIV=12,headings=big]{scrartcl}
\pdfoutput=1

\usepackage[a4paper, top=1.2in, bottom=1.2in, left=1.1in, right=1.1in]{geometry}
\usepackage{amsfonts}
\usepackage{amsmath,amssymb}
\usepackage{mathrsfs}
\usepackage[normal]{caption}

\usepackage{titlesec}
\titleformat*{\section}{\Large \sffamily \bfseries \boldmath}
\titleformat*{\subsection}{\large \sffamily \bfseries \boldmath}
\titleformat*{\subsubsection}{\sffamily \bfseries \boldmath}

\usepackage{cancel}
\usepackage[dvipsnames]{xcolor}
\usepackage{cite,bbm}
\usepackage{enumerate}
\usepackage{graphicx}
\graphicspath{{./Figs/}}
\usepackage{wrapfig}

\usepackage{bbold}

\renewenvironment{abstract}{%
\begin{minipage}{0.95\textwidth}
}
{\par\noindent\end{minipage}}

\usepackage[multiple]{footmisc}
\let\oldfootnote\footnote\renewcommand\footnote[1]{\oldfootnote{\hspace{2mm}#1}}

\definecolor{darkblue}{rgb}{0,0,0.9}
\usepackage{hyperref}
\hypersetup{
linktocpage,
colorlinks,
citecolor=darkblue,
filecolor=darkblue,
linkcolor=darkblue,
urlcolor=darkblue
}
\allowdisplaybreaks
\addtokomafont{disposition}{\rmfamily\boldmath}

\newcommand{\mc}{\mathcal}
\newcommand{\ms}{\mathscr}

\newcommand{\eff}{{\rm eff}}

\def\sla#1{\setbox0=\hbox{$#1$}\dimen0=\wd0
      \setbox1=\hbox{/} \dimen1=\wd1 \ifdim\dimen0>\dimen1
      \rlap{\hbox to \dimen0{\hfil/\hfil}} #1                        \else
      \rlap{\hbox to \dimen1{\hfil$#1$\hfil}}
      /   \fi}

\newcommand{\be}{\begin{equation}}
\newcommand{\ee}{\end{equation}}
\newcommand{\bea}{\begin{eqnarray}}
\newcommand{\eea}{\end{eqnarray}}

\newcommand{\nn}{\nonumber}

\DeclareOldFontCommand{\rm}{\normalfont\rmfamily}{\mathrm}
\DeclareOldFontCommand{\sf}{\normalfont\sffamily}{\mathsf}
\DeclareOldFontCommand{\tt}{\normalfont\ttfamily}{\mathtt}
\DeclareOldFontCommand{\bf}{\normalfont\bfseries}{\mathbf}
\DeclareOldFontCommand{\it}{\normalfont\itshape}{\mathit}
\DeclareOldFontCommand{\sl}{\normalfont\slshape}{\@nomath\sl}
\DeclareOldFontCommand{\sc}{\normalfont\scshape}{\@nomath\sc}

\begin{document}

%%%% TITLE PAGE
\begin{flushright}
\small
LAPTH-028/18
%CERN-PH-TH/2013-190
\end{flushright}
\vskip0.5cm

\begin{center}
%%%% TITLE
{\sffamily \bfseries \LARGE \boldmath
A gauged horizontal $SU(2)$ symmetry and $R_{K^{(\ast)}}$}\\[0.8 cm]
%%%% AUTHORS
{\normalsize \sffamily \bfseries Diego Guadagnoli$^a$, M\'eril Reboud$^a$ and Olcyr Sumensari$^b$} \\[0.5 cm]
\small
$^a${\em Laboratoire d'Annecy-le-Vieux de Physique Th\'eorique UMR5108\,, Universit\'e de Savoie Mont-Blanc et CNRS, B.P.~110, F-74941, Annecy Cedex, France}\\[0.1cm]
$^b${\em Dipartimento di Fisica e Astronomia `G. Galilei', Universit\`a di Padova, Italy \\ Istituto Nazionale Fisica Nucleare, Sezione di Padova, I-35131 Padova, Italy}\\
\end{center}

\medskip

\begin{abstract}
One of the greatest challenges for models of $b \to s$ anomalies is the necessity to produce a large contribution to a quark times a lepton current, $J_q \times J_\ell$, and to avoid accordingly large contributions to flavour-changing $J_q \times J_q$ and $J_\ell \times J_\ell$ amplitudes, which are severely constrained by data. We consider a gauged horizontal symmetry involving the two heaviest generations of all left-handed fermions. In the limit of degenerate masses for the horizontal bosons, and in the absence of mixing between the two heavier generations and the lighter one, such symmetry would make $J_q \times J_q$ and $J_\ell \times J_\ell$ amplitudes exactly flavour-diagonal. Mixing with the first generation is however inescapable due to the CKM matrix, and the above mechanism turns out to be challenged by constraints such as $D^0 -\bar D^0$ mixing. Nonetheless, we show that a simultaneous description of all data can be accomplished by simply allowing for non-degenerate masses for the horizontal bosons. Such scenario predicts modifications in several processes that can be tested at present and upcoming facilities. In particular, it implies a lower and upper bound for $\mc B (B \to K \mu^\pm \tau^\mp)$, an asymmetry between its two charge conjugated modes, and well-defined correlations with LFV in $\tau$ decays.
\end{abstract}

\vspace{0.8cm}

\renewcommand{\thefootnote}{\arabic{footnote}}
\setcounter{footnote}{0}
%%%% END TITLE PAGE

\section{Introduction}

\noindent
Data on $b \to s$ semi-leptonic transitions display persistent deviations with respect to Standard-Model (SM) predictions, hinting at a violation of lepton universality (LUV) \cite{Aaij:2014ora,Aaij:2017vbb}, namely at some new interaction that distinguishes among lepton species. Further hints of LUV exist for $b \to c$ semi-leptonic transitions \cite{Lees:2013uzd,Aaij:2015yra,Hirose:2016wfn,Aaij:2017deq}. At face value, the two sets of anomalies have similar significances of about 4$\sigma$ \cite{Amhis:2016xyh,Altmannshofer:2017yso,Capdevila:2017bsm,Ciuchini:2017mik,DAmico:2017mtc,Hiller:2017bzc,Geng:2017svp}, which justifies taking both datasets on an equal footing. From a theory point of view, such a stance is also motivated by the fact that the two sets of anomalies convey the same qualitative message (LUV), and that they concern currents that can be related by the SM $SU(2)_L$ symmetry, which is what one would expect of new effects arising above the electroweak symmetry-breaking (EWSB) scale \cite{Bhattacharya:2014wla,Alonso:2014csa,Buttazzo:2017ixm}. However, seeking an explanation of both sets of anomalies turns out to be problematic at the {\em quantitative} level, if nothing else because $b \to c$ and $b \to s$ data hint at $\sim 10-20$\% shifts in, respectively, a SM tree and a SM loop amplitude. If both shifts are to be explained through the same effective, $SU(2)_L$-invariant structure, its flavour-dependent coupling must come with a mechanism allowing for more suppressed effects in $b \to s$ than in $b \to c$. Numerous proposals in this direction have already been made in the literature, for example \cite{Bauer:2015knc}, that implements a tree- vs. loop-suppression mechanism akin to the SM one (on this model, see also \cite{Becirevic:2016oho,Cai:2017wry}), or a broken flavour symmetry whereby $b \to c \tau \nu$ and $b \to s \mu \mu$ effects arise as respectively first and third order in the breaking parameter \cite{Barbieri:2015yvd} (see also \cite{Buttazzo:2017ixm,Greljo:2015mma,Bordone:2017anc}). Many more such proposals exist in the literature, including a few UV-complete models \cite{Assad:2017iib,DiLuzio:2017vat,Calibbi:2017qbu,Bordone:2017bld,Barbieri:2017tuq,Becirevic:2018afm,Trifinopoulos:2018rna,Blanke:2018sro}. These models have to invariably withstand non-negligible constraints, in particular from certain low-energy precision observables \cite{Feruglio:2016gvd,Feruglio:2017rjo,Cornella:2018tfd,Feruglio:2018fxo,Hati:2018fzc} and/or direct searches \cite{Faroughy:2016osc,Altmannshofer:2017poe}. 
The overall challenge thus boils down to the difficulty of writing down a calculable model that explains a (coherent) set of LUV anomalies in $b \to s$ and $b \to c$ currents, with no observed departures in other directly related sets of data. Perhaps this difficulty is indicating that data are still premature to be taken quantitatively. 
In these circumstances, we choose to focus on $b \to s$ anomalies alone. In this paper we propose a mechanism for these discrepancies, that rests on two main requirements that data seem to convey: {\em (i)} the new dynamics explaining the $b \to s$ measurements must, directly or indirectly, involve the second and the third generation of quarks and leptons; {\em (ii)} it must yield large enough effects in the product of a quark times a charged-lepton bilinear, $J_q \times J_\ell$, and small enough effects elsewhere, in particular in flavour-changing $J_q \times J_q$ and $J_\ell \times J_\ell$ amplitudes.

\section{Model}
\newcommand{\GSM}{G_{\rm SM}}
\newcommand{\suh}{SU(2)_h}
Facts {\em (i)} and {\em (ii)} in the previous section suggest, for reasons that will be transparent shortly, the consideration of a `horizontal' group, with $SU(2)$ being the smallest one that may be at play.\footnote{References on the topic of horizontal symmetries for $B$-decay discrepancies include \cite{Crivellin:2015lwa,Alonso:2017bff,Cline:2017ihf}, but respective lines of arguments are quite distant from the one pursued here. In a context predating LUV in $B$ decays, the possibility of a fully gauged flavour group was discussed in Refs. \cite{Grinstein:2010ve,Guadagnoli:2011id}.} In this paper we invoke the possibility of a gauged $SU(2)$ horizontal symmetry. We consider the gauge group $\GSM \times G_h$, where $\GSM$ is the SM gauge group $SU(3)_c \times SU(2)_L \times U(1)_Y$, acting `vertically' in each generation, and $G_h = \suh$ is a horizontal group connecting the $2^{\rm nd}$ and $3^{\rm rd}$ generations as defined before EWSB. More generally, we may actually assume one $\suh$ symmetry for either chirality of fermions.\footnote{Our main line of argument was presented, in an entirely different context, in an old work by Cahn and Harari \cite{Cahn:1980kv}. See also \cite{Wilczek:1978xi}.} Anomaly cancellation is automatic for each chirality separately, and we will comment on it later.

In general, we can then augment the SM Lagrangian with the following terms
\be
\label{eq:dL_2x2}
\delta \ms L = \sum_{\ms F, a} \bar{\ms F} \left(g_L \gamma_\mu P_L G^{\mu\,a}_L +
g_R \gamma_\mu P_R G^{\mu\,a}_R \right) \tau^a {\ms F}~,
\ee
where $P_{L,R}$ are the usual chirality projectors, that we henceforth include in the gamma matrices for brevity, i.e.~$\gamma^\mu P_{L,R} = \gamma^\mu_{L,R}$. Furthermore, $\tau^a = \sigma^a/2$, and $G^{\mu\,a}_{L,R}$ are the gauge bosons of the horizontal symmetry for either chirality, whose masses are assumed to be larger than the EWSB scale. The fields $\ms F$ are such that
\be
\label{eq:F_2x2}
\ms F \equiv
\left(
\begin{array}{c}
f_2\\
f_3\\
\end{array}
\right)~,
\ee
with 2, 3 being generation indices in general not aligned with the mass eigenbasis. The symbol $f$ runs over all SM fermion species $u_{L,R}, d_{L,R}$, $\nu_L$, $\ell_{L,R}$.

An interesting phenomenological feature of the above interaction becomes apparent after integrating out the $G_L$ and $G_R$ gauge bosons.
One obtains the effective interactions
\be
\label{eq:dLeff}
\delta \ms L_{\rm eff} = - \sum_{\ms F, \ms F',a} 
\left\{ 
\frac{g_L^2}{2 M_{G_{La}}^2} \left( \bar{\ms F} \gamma^\mu_L \tau^a \ms F\right) \left(\bar{\ms F'} \gamma_{\mu\,L} \tau^a \ms F'\right) +
\frac{g_R^2}{2 M_{G_{Ra}}^2} \left( \bar{\ms F} \gamma^\mu_R \tau^a \ms F\right) \left(\bar{\ms F'} \gamma_{\mu\,R} \tau^a \ms F'\right)
\right\}~,
\ee
where both $\ms F$ and $\ms F'$ are defined as in Eq.~(\ref{eq:F_2x2}). Below the EWSB scale, where SM fermions acquire masses, the fields $\ms F$ undergo chiral unitary\footnote{\label{foot:2gen}For two generations, as in Eq.~(\ref{eq:F_2x2}), these transformations are actually not unitary. Here we are sacrificing accuracy for the sake of presenting the main argument. We will make notation more precise afterwards.} transformations of the kind
\be
\label{eq:F-rotation}
\ms F = \mc U_{\ms F} \hat{\ms F}
\ee
where $\hat{\ms F}$ denotes the mass-eigenbasis fields. After such transformations, the four-fermion structures generated by integrating out the $G_{La}$ horizontal bosons from Eq.~(\ref{eq:dLeff}) become
\be
\label{eq:dLeff_rotated}
\delta \ms L_{\rm eff} \propto \frac{1}{2 M_{G_{La}}^2} \left( \bar{\hat{\ms F}} \, \mc U^\dagger_{\ms F} \,\gamma^\mu_{L} \, \tau^a \, \mc U_{\ms F} \, \hat{\ms F}\right) \left(\bar{\hat{\ms F}}' \, \mc U^\dagger_{\ms F'} \, \gamma_{\mu\,L} \, \tau^a \, \mc U_{\ms F'} \,\hat{\ms F}'\right)~,
\ee
and analogous structures are generated by the $G_{Ra}$ terms. Hence, in either the left- or right-handed sector, these effective interactions have the form $\sum_a J_{\ms F}^a \times J_{\ms F'}^a / M^2_{Ga}$. Then, from Eq.~(\ref{eq:dLeff_rotated}) one sees that, {\em in the limit of mass degeneracy} across the horizontal bosons of either sector, products of currents involving the same fermions, $\ms F = \ms F'$ (for example, both equal to down-type quarks), are such that the rotations $\mc U_{\ms F}$ can be shuffled in the definition of the horizontal gauge-boson fields, and {\em the corresponding fermion bilinears can be taken as flavour diagonal in all generality \cite{Cahn:1980kv}}.

As advertised above, this property is welcome, because new off-diagonal contributions to products of currents with $\ms F = \ms F'$ are severely constrained by data, in particular meson mixings ($\ms F = \ms F' = d$) and respectively purely leptonic LFV decays such as $\tau \to 3 \mu$ ($\ms F = \ms F' = \ell$). In the absence of such mechanism, these processes would pose the most formidable constraints, as they do in other SM extensions by gauge groups, in particular $W', Z'$ ones (see in particular \cite{Greljo:2015mma,Boucenna:2016wpr,Buras:2013dea,Descotes-Genon:2017ptp}). In fact, these constraints are perhaps the most outstanding reason in favour of models with lepto-quark mediators, whereby $J_\ell \times J_\ell$ and $J_q \times J_q$ currents\footnote{At least in lepto-quark models where di-quark couplings are absent \cite{Dorsner:2016wpm}.} are generated only at loop level.

There is actually a subtlety, already mentioned in footnote \ref{foot:2gen}. Although above the EWSB scale $\suh$ involves (by construction) only the $2^{\rm nd}$ and $3^{\rm rd}$ generations, mixing beneath this scale involves all the three generations. As a consequence, the $\mc U_{\ms F}$ matrices in Eq.~(\ref{eq:F-rotation}) are not exactly unitary, implying that the contributions to processes such as meson mixings, as well as to decays involving only leptons, are non-zero. It is true that these contributions will be parametrically suppressed by powers of the mixing between the two heavier and the light generation. However, a non-zero mixing onto the $1^{\rm st}$ generation translates into contributions to light-fermion processes like $K^0 - \bar K^0$ mixing and $\mu \to 3 e$ for example, which are well-known to be very constraining \cite{Isidori:2010kg}. We will discuss such effects in detail in the analysis.

We need meanwhile to generalize the formalism in Eqs. (\ref{eq:dL_2x2})-(\ref{eq:dLeff_rotated}) to account for three-generation mixing. We then define
\be
\label{eq:F}
\ms F \equiv
\left(
\begin{array}{c}
f_1\\
f_2\\
f_3\\
\end{array}
\right)~.
\ee
The Lagrangian shift in Eq.~(\ref{eq:dL_2x2}) becomes
\be
\label{eq:dL}
\delta \ms L = \sum_{\ms F} \bar{\ms F} \left(g_L \gamma_{\mu \, L} G^{\mu\,a}_L +
g_R \gamma_{\mu \, R} G^{\mu\,a}_R \right) T^a {\ms F}~,
\ee
with namely the replacement $\tau^a \to T^a$, where 
\be
\label{eq:Ta}
T^a \equiv \left(
\begin{array}{cc}
0_{1 \times 1} & \\
& \tau^a\\
\end{array}
\right)~,
\ee
and Eqs. (\ref{eq:dLeff}) and (\ref{eq:dLeff_rotated}) will change accordingly. It is this Lagrangian that we will use in the analysis.

We reiterate that the argument leading to flavour-diagonal $J_q \times J_q$ and $J_\ell \times J_\ell$ amplitudes holds for horizontal bosons with degenerate masses, which needn't be the case. In fact, a departure from the hypothesis of exact mass degeneracy will be instrumental for our framework to withstand the constraints from, in particular, $D^0 - \bar D^0$ mixing. Effects on this and other observables will thus be parametric in the mass splittings among the horizontal bosons. As concerns the performance of Eq.~(\ref{eq:dL}) in explaining $R_{K^{(*)}}$, we note that the introduction of a sizeable contribution to the quark right-handed bilinear is expected to upset the relation $R_{K^*} \simeq R_K$ \cite{Hiller:2014ula}. The bulk of our analysis will therefore assume $g_R = 0$.

Before concluding, two points deserve discussion. First, the question whether the enlarged gauge group may introduce anomalies, even with just the SM matter content. A simple horizontal gauge group $G_h$ introduces two potentially worrisome anomaly diagrams: the one with $G_h^3$ and the one with $G_h^2 \times U(1)_Y$. Having chosen $G_h$ to be an $SU(2)$ group, the first diagram vanishes -- whereas it wouldn't for larger simple groups such as $SU(3)$. As concerns the second diagram, it likewise vanishes, and it does so for the same reason also at work within the SM, namely the rather magical compensation of the quark vs. lepton $U(1)_Y$ quantum numbers of either chirality. So this anomaly cancels separately for an $\suh$ coupled only to left-handed fermions or to right-handed ones.

A second outstanding question concerns the specification of the Yukawa sector, addressing for example how a horizontal symmetry involving the two heavier generations may be compatible with the observed fermion masses and mixing. A first possible scenario would be to invoke a Yukawa sector bilinear in SM fermion fields. Such scenario requires the introduction of new scalar representations that are charged under $G_h$, notably in Yukawa terms with products between 1$^{\rm st}$- and 2$^{\rm nd}$- or 3$^{\rm rd}$-generation SM fermions. The resulting construction resembles general models with extended Higgs sectors, that will reintroduce tree-level contributions to the very FCNC processes that our initial symmetry argument is designed to guard against. A more promising possibility is to consider a scenario akin to partial compositeness \cite{Kaplan:1991dc}, where the UV Yukawa terms involve the product between SM fermions, new vector-like fermions $\Psi$ as well as suitable scalar representations $\Phi_i$ to break $SU(2)_h$ spontaneously. For definiteness, one could consider the following field content
\be
\label{eq:new_fields}
\Psi^U_{L,R} \sim ({\bf 3}, {\bf 1}, 2/3; {\bf 2})~,~~~ \Psi^D_{L,R} \sim ({\bf 3}, {\bf 1}, -1/3; {\bf 2})~,~~~\Phi_{1,2} \sim ({\bf 1}, {\bf 1}, 0; {\bf 2})~,
\ee
where transformation properties refer to $G_{\rm SM}\times G_h$. This field content gives rise to the following renormalizable Lagrangian terms
\bea
\delta \ms L_Y &\supset& m_U \bar \Psi^U_L \Psi^U_R ~+~ m_D \bar \Psi^D_L \Psi^D_R \nn \\
&+& \sum_{a=1,2}\sum_{i = 1, ..., 3} \Bigl( (Y_U)_{ai} \overline {\Psi^U_L} \Phi_a (u_{R})_i + (Y_D)_{ai} \overline {\Psi^D_L} \Phi_a (d_{R})_i \Bigl) \nn \\
&+& c_U \bar F_Q \tilde H \Psi_R^U + c_D \bar F_Q H \Psi_R^D ~+~ {\rm H.c.}~,
\eea
where $i$ is a flavour index, $F_Q = ((Q_L)_2, (Q_L)_3)^T$, with $_{2,3}$ generation indices, and $H$ is the SM Higgs doublet. SM Yukawa terms for quarks would then arise after integrating out the heavy $\Psi$ and assigning vev's to the $\Phi$'s. Two scalar fields are needed in Eq. (\ref{eq:new_fields}) in order to generate rank-3 effective Yukawa matrices. The number of dimensionless parameters and mass scales thus involved is sufficient to accommodate quark masses and mixing. An entirely similar construction allows to also address lepton masses and mixing. Gauge models for $b \to s$ anomalies implementing a similar mechanism are Refs. \cite{Altmannshofer:2014cfa,Sierra:2015fma,Crivellin:2015mga}.

\subsection{Low-energy theory and notation} \label{sec:th}

The most general dimension-six Hamiltonian describing the transition $b \to s \ell_1^-\ell_2^+$ with $\ell =e,\mu,\tau$ reads~(we adopt the normalization in \cite{Bobeth:2001sq})
\be
\label{eq:hamiltonian-bsll}
\begin{split}
  \mc H_{{\rm eff}} = -\frac{4
    G_F}{\sqrt{2}}V_{tb}V_{ts}^* &\Bigg{\lbrace} \sum_{i=1}^6
  C_i(\mu)\mc O_i(\mu)+\sum_{i=7,8}
  \Big{[}C_i(\mu) \mc O_i(\mu)+\left(C_{i}(\mu)\right)^\prime \left(\mc O_{i}(\mu)\right)^\prime\Big{]}\\
& + \sum_{i=9,10,S,P}
  \Big{[} C^{\ell_1 \ell_2}_i(\mu)\mc O^{\ell_1 \ell_2}_i(\mu) + \left(C^{\ell_1 \ell_2}_{i}(\mu)\right)^\prime \left(\mc O^{\ell_1 \ell_2}_{i}(\mu)\right)^\prime\Big{]}\Bigg{\rbrace}
+{\rm h.c.},
\end{split}
\ee
where $C_i(\mu)$ and $C_i^{\ell_1 \ell_2}(\mu)$ are the Wilson coefficients, while the effective operators relevant to our study are defined by
\be
\label{eq:C_LFV}
\begin{split}
\mc O_{9}^{\ell_1\ell_2}
  =\frac{e^2}{(4\pi)^2}(\bar{s}\gamma_{\mu \, L} b)(\bar{\ell}_1\gamma^\mu\ell_2)~,~~~~
\mc O_{10}^{\ell_1\ell_2} 
  =\frac{e^2}{(4\pi)^2}(\bar{s}\gamma_{\mu \, L} b)(\bar{\ell}_1\gamma^\mu \gamma^5\ell_2)~,
\end{split}
\ee
in addition to the electromagnetic-dipole operator $\mc O_7=e/(4\pi)^2m_b (\bar{s}\sigma_{\mu\nu}P_R b)F^{\mu\nu}$. The chirality-flipped operators $\mc O_i^\prime$ are obtained from $\mc O_i$ by the replacement $P_L\leftrightarrow P_R$. Henceforth we will split Wilson coefficients as $C_i = C_i^{\rm SM} + \delta C_i$ so that the SM limit \cite{Buras:1994dj,Misiak:1992bc,Buchalla:1995vs} for the Wilson coefficients is obtained by $\delta C_i^{(\prime)} = 0$. Using the Hamiltonian given above, it is straightforward to compute the decay rates of $B_s\to \ell_1^-\ell_2^+$, $B\to K^{(\ast)}\ell_1^-\ell_2^+$ \cite{Gratrex:2015hna,Becirevic:2016zri}, and other similar modes, including radiative ones, given that the effective Hamiltonian is the same \cite{Guadagnoli:2016erb,Guadagnoli:2017quo}.

The pattern of observed departures of {\em all} $b \to s \ell \ell$ data from the SM finds a straightforward interpretation within the above EFT framework \cite{Altmannshofer:2017yso,Capdevila:2017bsm,Ciuchini:2017mik,DAmico:2017mtc,Hiller:2017bzc,Geng:2017svp}. Interestingly, among the preferred operators to explain the anomalies is the product of two left-handed currents \cite{Hiller:2014yaa,Ghosh:2014awa}, i.e. $\delta C_9^{\mu \mu} = - \delta C_{10}^{\mu \mu}$. Such solution is especially appealing theoretically, as it can naturally be expressed in terms of the $SU(2)_L$-invariant fields $Q_L$ and $L_L$ \cite{Bhattacharya:2014wla,Alonso:2014csa}, which is what one would expect of new effects generated above the electroweak symmetry-breaking (EWSB) scale. In order to obtain a `conservative' estimate of the allowed range for these Wilson coefficients, we use the results quoted in Ref.~\cite{DAmico:2017mtc} using only the ratios $R_K$ and $R_{K^\ast}$, as well as the leptonic decay $\mc B(B_s\to \mu^+\mu^-)$. The result of the fit to $1\sigma$ accuracy is
\be
\label{eq:C9value}
\delta C_9^{(\mu)} = - \delta C_{10}^{(\mu)} \in [-0.85, -0.50]~.
\ee

\section{Scenario 0} \label{sec:scenario0}

We then start from the effective Lagrangian Eq.~(\ref{eq:dLeff}), keeping henceforth only the left-handed interaction as motivated in the Introduction, and with mass-degenerate horizontal bosons. In order to simplify notation, we will in the rest of our paper remove the $L$ subscript from the $G_{La}$ fields (denoted simply as $G_a$) as well as from the quark and lepton multiplets in generation space. We will instead keep the subscript in $g_L$, in order that this coupling not be confused with SM ones.

Fermionic fields $\ms F$, Eq.~(\ref{eq:F}), are rotated to the mass eigenbasis through chiral transformations $\mc U_{\ms F}$. The most general parameterization compatible with the $\suh$ symmetry would be of the form
\be
\label{eq:UF=block}
\mc U_{\ms F} = {\rm diag}(\exp(i \phi_{\ms F}), \exp(i \Phi_{\ms F}) \Sigma_{\ms F})~,
\ee
where $\Sigma_{\ms F} = \exp(i \vec \theta_{\ms F} \, \vec \sigma / 2)$ parameterizes a general $SU(2)$ transformation acting on the $f_{2,3}$ components of the fermion ${\ms F}$, see Eq.~(\ref{eq:F}). The dependence on the phases $\phi_{\ms F}$ and $\Phi_{\ms F}$ cancels in quark and lepton bilinears. The dependence on the $SU(2)$ transformation in turn cancels in effective-Lagrangian terms involving one single fermion species, i.e.~terms of the form in Eq.~(\ref{eq:dLeff_rotated}) with $\ms F = \ms F'$, because of the argument made below that equation. As noted there, this mechanism \cite{Cahn:1980kv} makes $J_q \times J_q$ and $J_\ell \times J_\ell$ amplitudes flavour diagonal (at tree level), thus preventing dangerous contributions to processes such as $B_s - \bar B_s$ mixing and $\tau \to 3 \mu$.

Let us focus on the interaction term involving down-type quarks and charged leptons, $\delta \ms L_{\eff}^{DL}$, which is of direct interest to us. Eq.~(\ref{eq:dLeff}) implies
\be
\label{eq:dLeffDL}
\delta \ms L_{\eff}^{DL} = - \dfrac{g_L^2}{m_G^2} \Bigg[\bar{\hat D} \Big( \mc U_D^\dagger\, \gamma^\mu_L \, T^a \, \mc U_D \Big) \hat D\Bigg]  \Bigg[\bar{\hat L} \Big(\mc U_L^\dagger \,\gamma_{\mu \,L} \, T^a \,\mc U_L \Big{)}\hat L\Bigg{]}~,
\ee
where $\hat D=(d_L, s_L, b_L)^T$ and $\hat L=(e_L, \mu_L, \tau_L)^T$ denote left-handed mass-eigenstate fermions and $m_G$ is the common mass of the $G_a$ horizontal bosons. Using the argument below Eq.~(\ref{eq:dLeff_rotated}), one can rewrite Eq.~(\ref{eq:dLeffDL}) as
\be
\label{eq:dLeffDL_rewritten}
\delta \ms L_{\eff}^{DL} = - \dfrac{g_L^2}{m_G^2} \Bigg[\bar{\hat D} \Big( \gamma^\mu_L \, T^a \Big) \hat D\Bigg]  \Bigg[\bar{\hat L} \Big(\mc U_{DL}^\dagger \, \gamma_{\mu \, L} \, T^a \, \mc U_{DL} \Big{)}\hat L\Bigg{]}~,
\ee
where $\mc U_{DL} \equiv \mc U_D^\dagger \, \mc U_L$ can be parameterized as in Eq.~(\ref{eq:UF=block}). Again, phase terms disappear in the lepton bilinear. As concerns the $SU(2)$ transformation $\Sigma_{DL}$, it is convenient to express it in terms of Euler angles:
\be
\label{eq:Sigma}
\Sigma_{DL} = \exp(-i \alpha_{DL} \sigma_3) \exp(-i \beta_{DL} \sigma_2) \exp(-i \gamma_{DL} \sigma_3)~.
\ee
Neglecting neutrino masses, the $\gamma_{DL}$ phase can be absorbed in the definition of the charged-lepton fields. A similar rephasing of the $D$ component fields to absorb $\alpha_{DL}$ would shuffle this phase to the CKM matrix, so $\alpha_{DL}$ and $\beta_{DL}$ are physical.\footnote{Note that, if one sets $\alpha_{DL} = 0$, Eq.~(\ref{eq:Sigma}) amounts to a generic rotation between the second and third generations. One sees that in the small-$\beta_{DL}$ limit, operators that violate generation number (either of the quark or lepton one, or both) by one or two units are proportional to one or two powers of $\beta_{DL}$, respectively, whereas only operators that conserve generation numbers are present in the limit of no mixing \cite{Cahn:1980kv}.}

\begin{figure}[t]
\centering
	\includegraphics[width=0.49\linewidth]{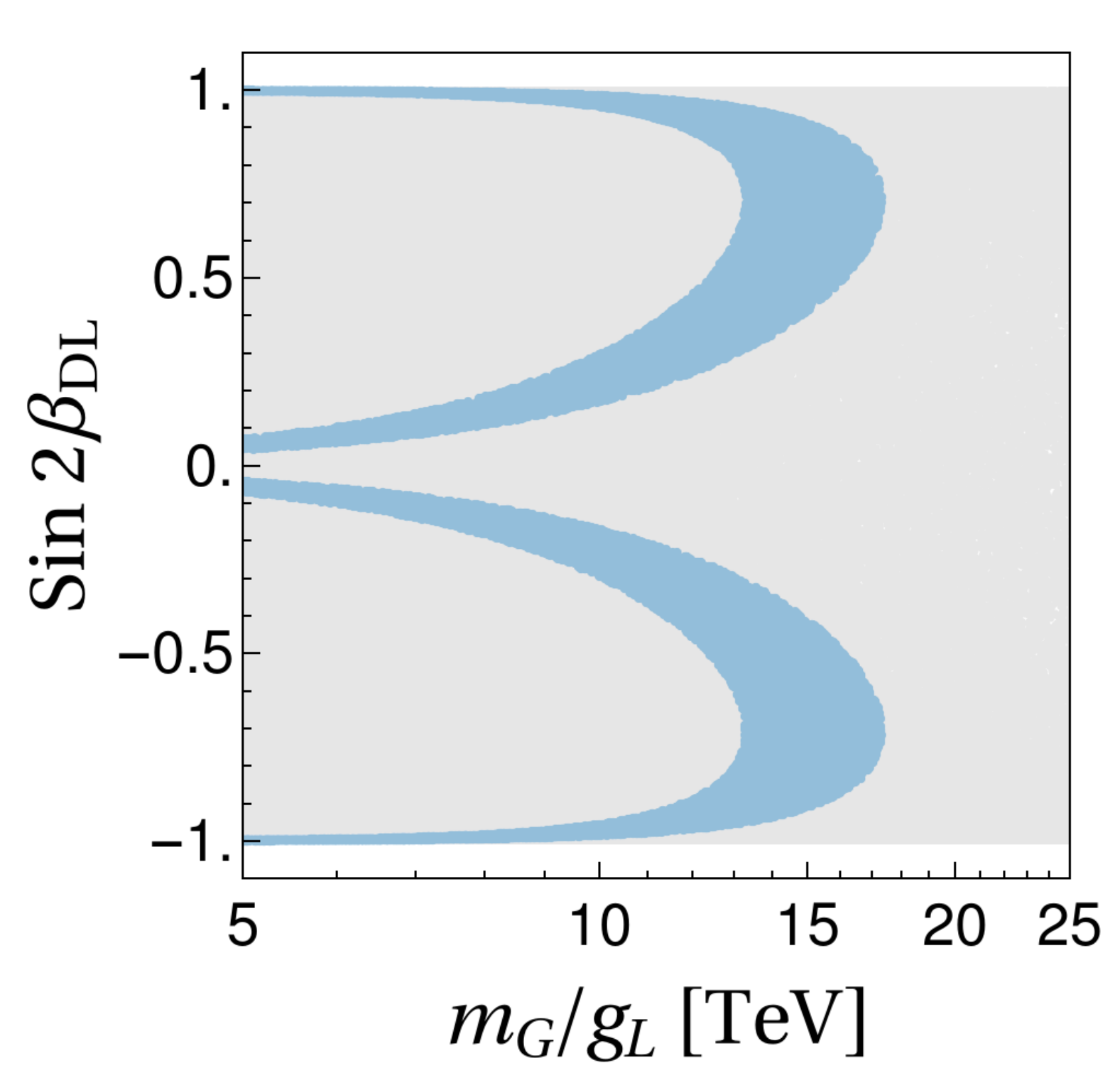}
	\hfill
	\includegraphics[width=0.49\linewidth]{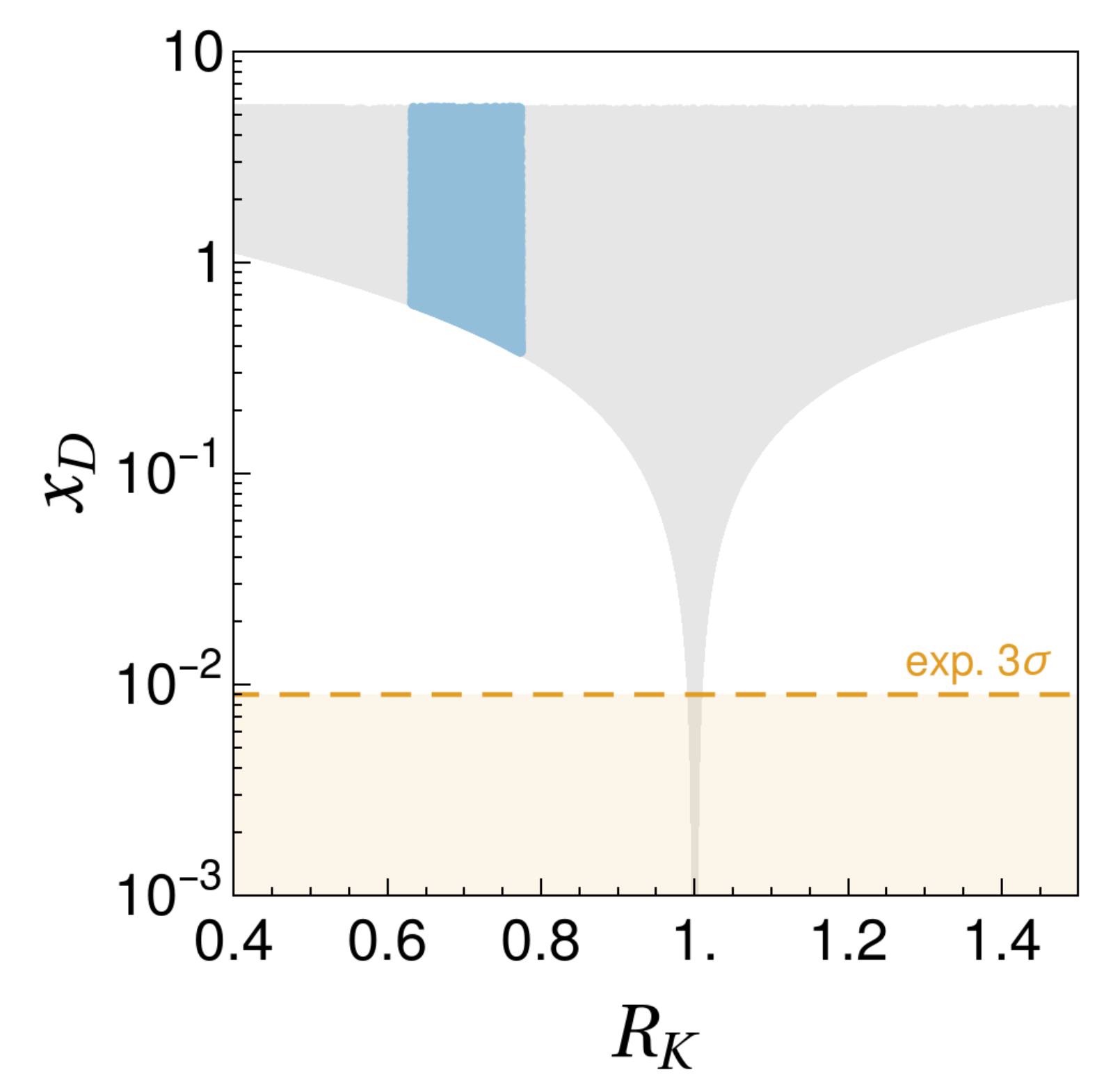}
\caption{\small \sl Two relevant projections of the model parameter space within scenario 0. Gray points fulfil all constraints except $R_K$ and $x_D$. Light-blue points fulfil also $R_K$. The dashed line represents the $3\sigma$ upper bound on $x_D$ from Ref. \cite{Amhis:2016xyh}.}
\label{fig:scenario0}
\end{figure}

We next discuss the most general effects within the parameterization in Eq.~(\ref{eq:Sigma}), in particular whether scenario 0 explains $R_{K^{(*)}}$ and what are the relevant constraints. By matching the Lagrangian in Eq.~(\ref{eq:dLeffDL_rewritten}) with Eq.~\eqref{eq:hamiltonian-bsll}, we obtain the following Wilson-coefficient shifts $\delta C_9^{ij}$
\begin{align}
\label{eq:wc-mumu}
\delta C_9^{\mu\mu} = - \delta C_9^{\tau\tau} &= e^{-2i \alpha_{DL}} \, \frac{\sin 2 \beta_{DL} \, g_L^2}{8 m_G^2} \cdot \frac{2\pi v^2}{\alpha_{{\rm em}} V_{tb} V_{ts}^\ast}\,, \\
\label{eq:wc-taumu}
\delta C_9^{\tau\mu} &= e^{-2i \alpha_{DL}} \, \frac{\cos^2\beta_{DL} \, g_L^2}{4 m_G^2} \cdot \frac{2\pi v^2}{\alpha_{{\rm em}} V_{tb} V_{ts}^\ast} \,,\\ 
\label{eq:wc-mutau}
\delta C_9^{\mu\tau} &= - e^{-2i \alpha_{DL}} \, \frac{\sin^2 \beta_{DL} \, g_L^2}{4 m_G^2} \cdot \frac{2\pi v^2}{\alpha_{{\rm em}} V_{tb} V_{ts}^\ast}\,,
\end{align}
and one has also $\delta C_{10}^{ij} = - \delta C_{9}^{ij}$. The last factor on the r.h.s. of either of Eqs. (\ref{eq:wc-mumu})-(\ref{eq:wc-mutau}) corresponds to an effective scale $\approx 34$ TeV \cite{DAmico:2017mtc}. From Eqs. (\ref{eq:wc-mumu})-(\ref{eq:wc-mutau}) we also see that largest (in magnitude) shifts to the considered Wilson coefficients are obtained for $\alpha_{DL} \approx 0$ because of the nearly real SM normalization in the usual CKM conventions. The bounds in Eq.~(\ref{eq:C9value}) then would amount to
\be
\label{eq:RK-constraint}
\dfrac{m_G^2}{\sin 2 \beta_{DL} \, g_L^2} \in ( [13, 17]~{\rm TeV} )^2~,
\ee
and the correlation between the rotation $\beta_{DL}$ and the scale $m_G / g_L$ is shown in the left panel of Fig.~\ref{fig:scenario0}. Eqs. (\ref{eq:wc-mumu})-(\ref{eq:wc-mutau}) instruct us that a nonzero value of $\beta_{DL}$ is needed to successfully explain $R_{K^{(\ast)}}$. This requirement automatically implies a nonzero value of $\delta C_9^{\tau\mu}$ and $\delta C_9^{\mu\tau}$. Therefore, LFV in the $b\to s \mu^\pm\tau^\mp$ channel is a very distinctive signature of our scenario, that we will discuss more extensively in sec.~\ref{sec:LFV}. It is likewise noteworthy that our framework predicts an asymmetry between $b\to s \mu^-\tau^+$ and $b\to s \mu^+\tau^-$, controlled by the $\beta_{DL}$ value.

There is, however, an important caveat around Eq.~(\ref{eq:UF=block}). Below the EWSB scale, the rotations $\mc U_{D}$ and $\mc U_{U}$ cannot both have the form in Eq.~(\ref{eq:UF=block}), because $\mc U_{U}^\dagger \, \mc U_{D} = V_{\rm CKM}$. This implies that, although the $\suh$ symmetry prevents the occurrence of flavour-violating $J_q \times J_q$ and $J_\ell \times J_\ell$ effects for scales above the EWSB one, $J_q \times J_q$ effects will be induced for lower scales, because of $V_{\rm CKM}$-induced mixing.\footnote{On the other hand, $J_\ell \times J_\ell$ effect will remain tiny because of the very small neutrino masses.} The most constraining of these effects turns out to be the mass difference in the $D^0 - \bar D^0$ system, $\Delta M_D$. We imposed that the latest global fit to the related parameter $x_D$ \cite{Amhis:2016xyh} be saturated by our model's short-distance prediction for the same quantity, that we estimated using Ref. \cite{Golowich:2007ka}.\footnote{For all details on the implementation, see Ref. \cite{Guadagnoli:2010sd}.} We believe that this approach be justified, given that the possible range for the SM contribution to $\Delta M_D$ encompasses several orders of magnitude \cite{Falk:2004wg}.

At face value,\footnote{I.e.~barring a tuning of order $10^{-2}$ between the SM and the new-physics contribution.} and quite unexpectedly, the $x_D$ constraint excludes our scenario 0. However our model is suited for straightforward, and actually plausible, generalizations. The latter fall in at least two categories: {\em (i)} mass splittings among the three vector bosons of the $\suh$ symmetry; {\em (ii)} (small) mixing terms between the first and the two heavier generations in the $T^a$ matrices of Eq.~(\ref{eq:Ta}). As we will discuss in the rest of the paper, the first generalization turns out to be sufficient to pass all constraints.

\section{Scenario 1} \label{sec:scenario1}

The most straightforward generalization of the scenario discussed so far is to allow for non-degenerate masses for the gauge bosons of the $\suh$ symmetry.
The simplest mass splitting is such that two masses stay degenerate. Such splitting can be achieved via the symmetry-breaking pattern advocated in Ref. \cite{Monich:1980rr}, i.e.~through one spin-$1/2$ {\em and} one spin-1 fundamental scalar representation.

With split $G_{a}$ masses, we are no more allowed to bundle the two unitary transformations $\mc U_D$ and $\mc U_L$ in one single transformation, as in Eq.~(\ref{eq:dLeffDL_rewritten}). Our effective interaction is thus
\be
\label{eq:dLeffDL^a}
\delta \ms L_{\eff}^{DL} = - \sum_a \dfrac{g_L^2}{m_{G_a}^2} \Bigg[\bar{\hat D} \Big( \mc U_D^\dagger\, \gamma^\mu_L \, T^a \, \mc U_D \Big) \hat D\Bigg]  \Bigg[\bar{\hat L} \Big(\mc U_L^\dagger \,\gamma_{\mu \, L} \, T^a \,\mc U_L \Big{)}\hat L\Bigg{]}~,
\ee
i.e.~akin to Eq.~(\ref{eq:dLeffDL}) but for non-degenerate $G_{a}$ masses. 
In a notation straightforwardly generalizing that in Eq.~(\ref{eq:Sigma}), the matrices $\mc U_{D,L}$ will introduce the rotations $\beta_D$ and $\beta_L$, as well as the phase parameters $\alpha_{D,L}$ and $\gamma_{D,L}$. To the extent that we do not consider $CP$-violating observables, non-zero values for these phase terms serve only to suppress the magnitude of the Wilson coefficients relevant to our analysis -- see discussion around Eqs. (\ref{eq:wc-mumu})-(\ref{eq:wc-mutau}). We will therefore set $\alpha_{D,L} = \gamma_{D,L} = 0$ and focus on the rotations $\beta_D$ and $\beta_L$.
We note that, after EWSB, the above choice for the $\mc U_D$ matrix allows to subsequently set $\mc U_U = \mc U_D V_{\rm CKM}^\dagger$. An important assumption concerns then the $\mc U_D$ parameterization. We assume, on phenomenological grounds, that $\mc U_D$ still fulfils the block-diagonal form in Eq. (\ref{eq:Sigma}), because this guarantees  the absence of tree-level effects in $K^0 - \bar K^0$ mixing.
%The fact that $\mc U_D$ fulfils the form in Eq. (\ref{eq:Sigma}) guarantees the absence of effects in $K^0 - \bar K^0$ mixing.

Quite remarkably a scenario with
\be
\label{eq:parameters_scenario1}
m_{G_{1}} = m_{G_{2}} \ll m_{G_{3}} ~~~~\&~~~~ |\sin2\beta_D| \ll |\sin2\beta_L|
\ee
accounts at one stroke for new effects in $b \to s \mu \mu$ as large as measured and is compatible with the SM-like results in all other collider datasets. We remark from the outset that, allowing for a mass hierarchy between the horizontal gauge bosons amounts to completely forsaking the argument made below Eq. (\ref{eq:dLeff_rotated}). This makes scenarios 0 and 1 completely different at the level of the underlying mechanisms. Within scenario 0 (degenerate horizontal-boson masses) the flavour diagonality of $J_q \times J_q$ and $J_\ell \times J_\ell$ currents would be the result of an underlying global symmetry coming with the postulated $\suh$ group. However, off-diagonalities are inescapable because of the CKM matrix, and the result is too large a contribution to $D^0 - \bar D^0$ mixing. Within scenario 1, one allows for non-degenerate masses, and phenomenological viability chooses a hierarchical pattern, i.e.~one of O(1) breaking of the mentioned {\em global} symmetry.

The basic mechanism at work can be straightforwardly understood by inspection of the model's prediction of $D^0 - \bar D^0$ mixing and $R_K$, that scenario 0 fell short to describe simultaneously. We will see that, with these two phenomenological requirements fulfilled, all other constraints fall in place, either because of the pattern in Eq. (\ref{eq:parameters_scenario1}), or because of the underlying $\suh$ symmetry. We will next discuss all these requirements in turn.

\begin{figure}[ht!]
\centering
	\includegraphics[width=0.475\linewidth]{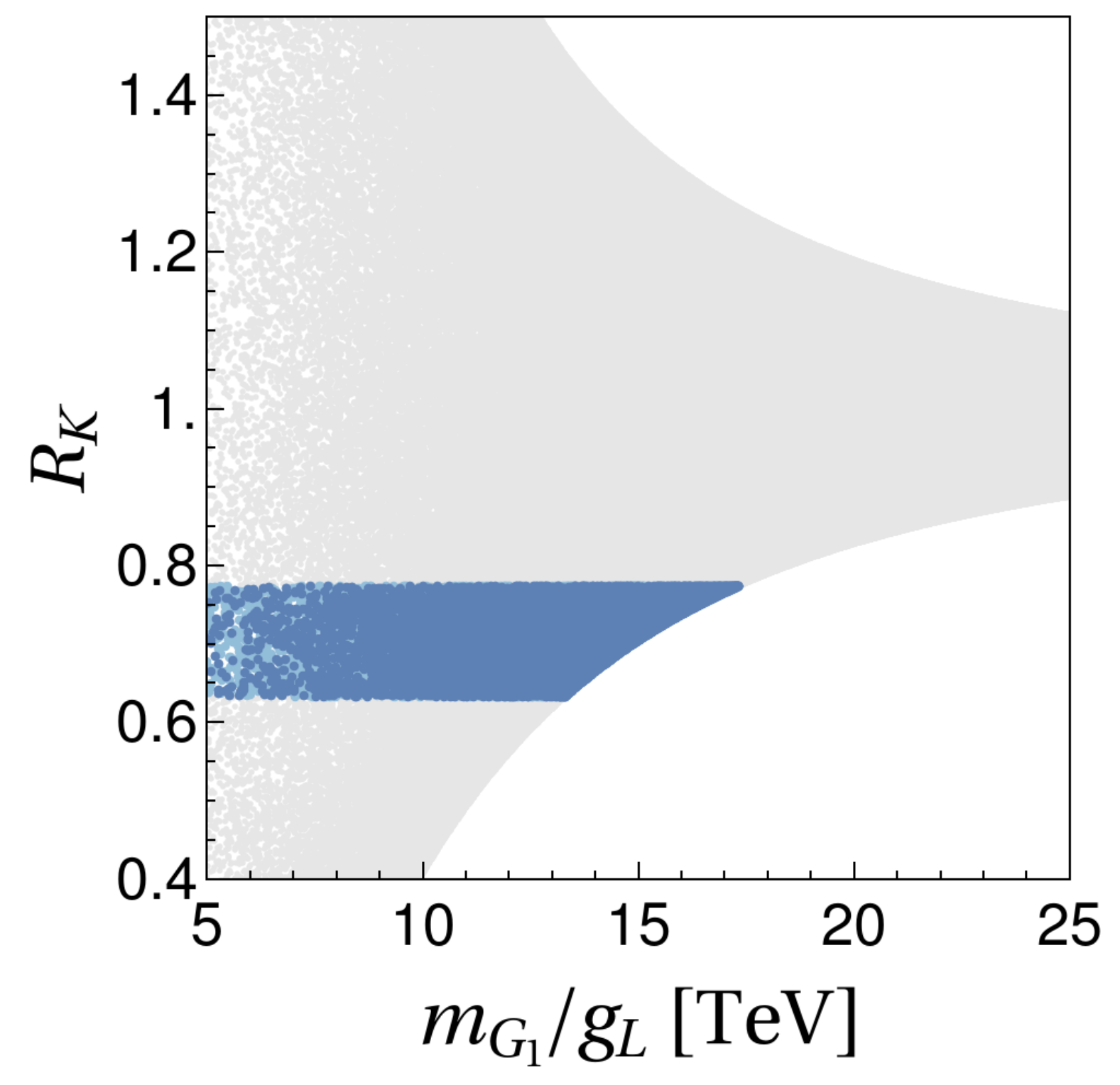}
	\hfill
	\includegraphics[width=0.500\linewidth]{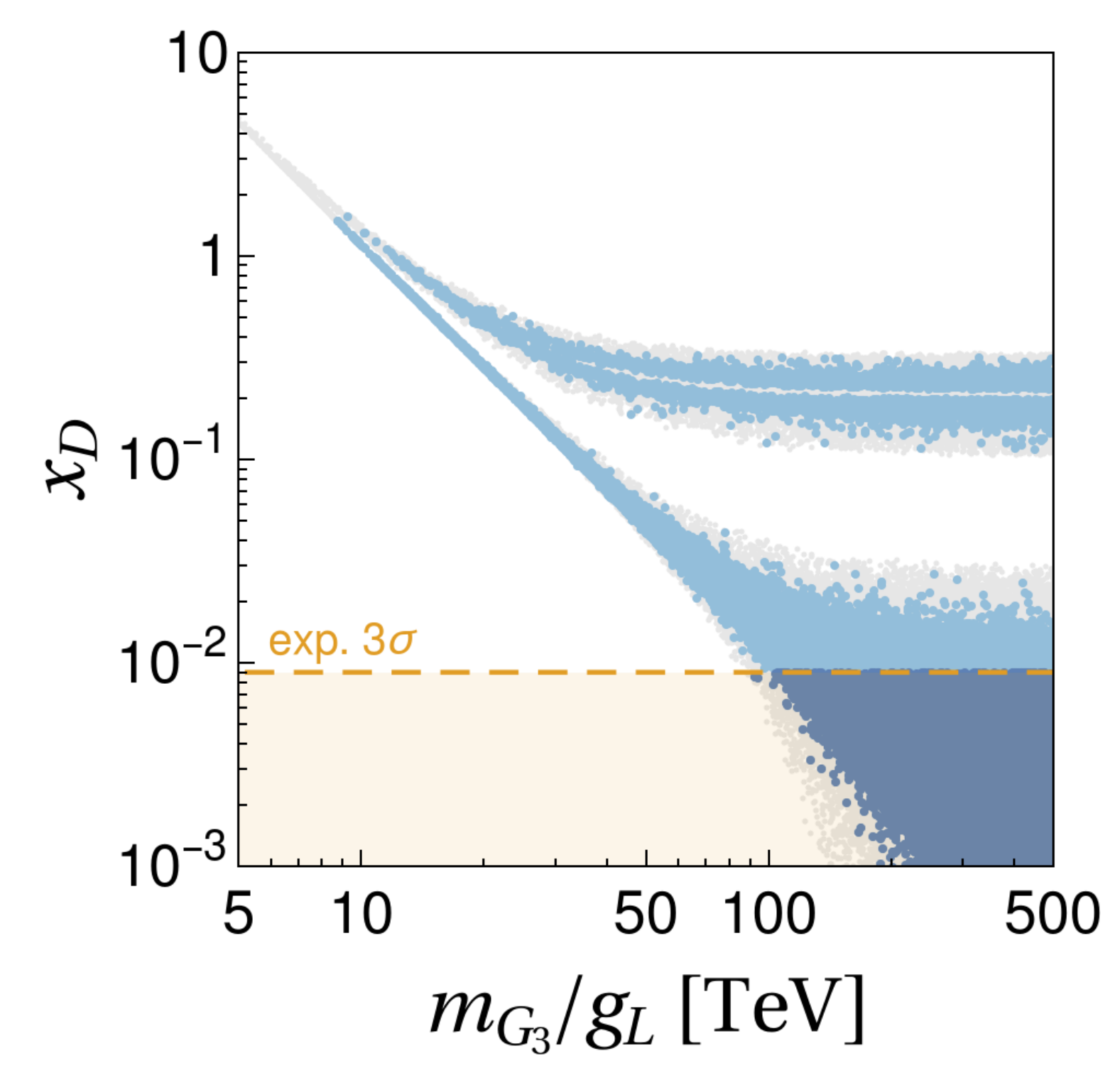}
\caption{\small \sl Color code henceforth: grey points fulfil all constraints except $R_K$ and $x_D$; light-blue points also fulfil $R_K$, but not $x_D$; dark-blue points fulfil all constraints. Left panel: $R_K$ vs. $m_{G_{1}}/g_L$ as implied by Eq.~(\ref{eq:wc-mumu-scenario1}). Right panel: effective mass scale for the heavier among the $G_{a}$ bosons, as required by the $x_D$ constraint. The horizontal dashed line denotes the $3\sigma$ upper bound on $x_D$ \cite{Amhis:2016xyh}.
}
\label{fig:scenario1}
\end{figure}

\begin{figure}[ht!]
\centering
	\includegraphics[width=0.49\linewidth]{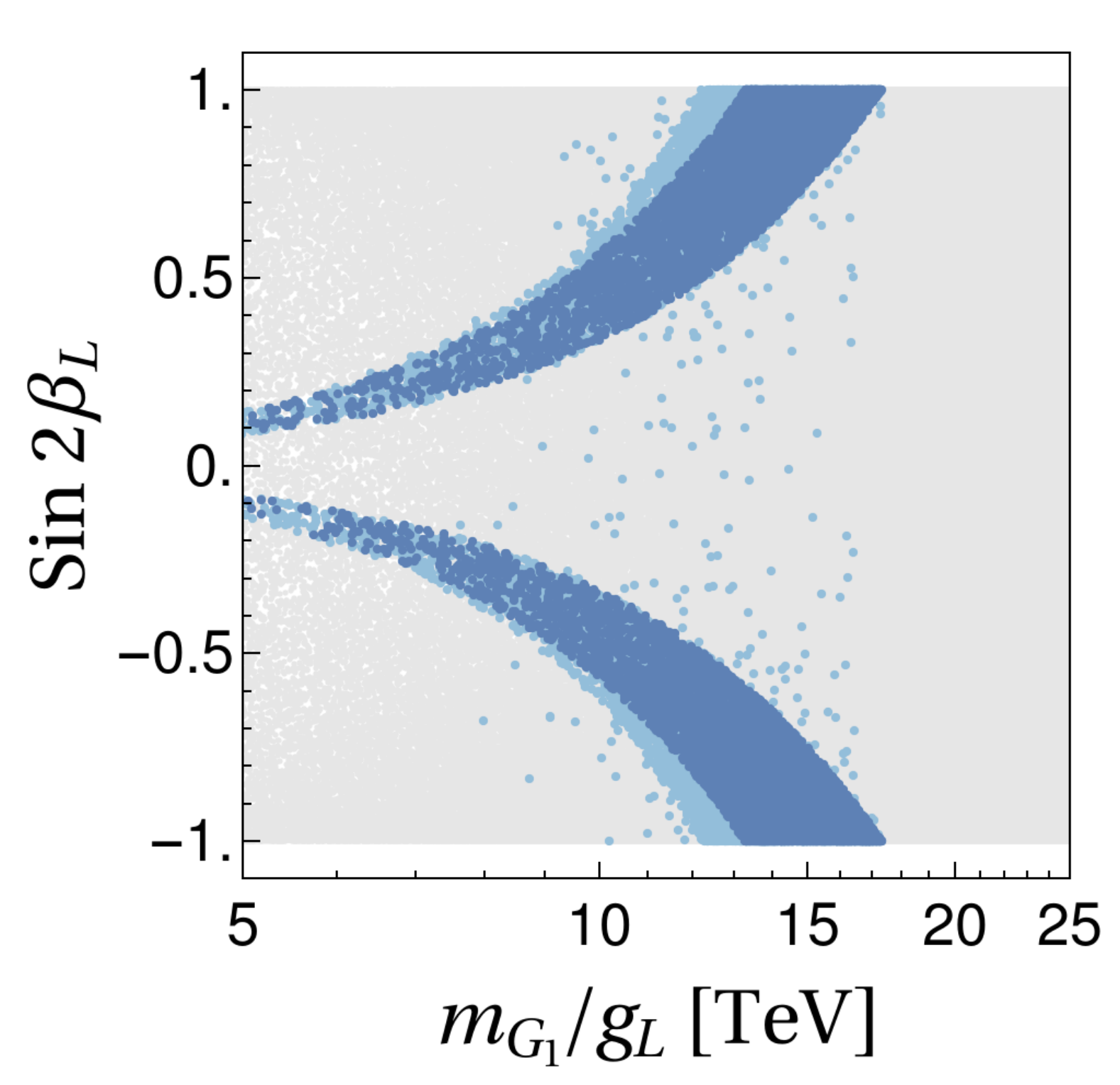}
	\hfill
	\includegraphics[width=0.49\linewidth]{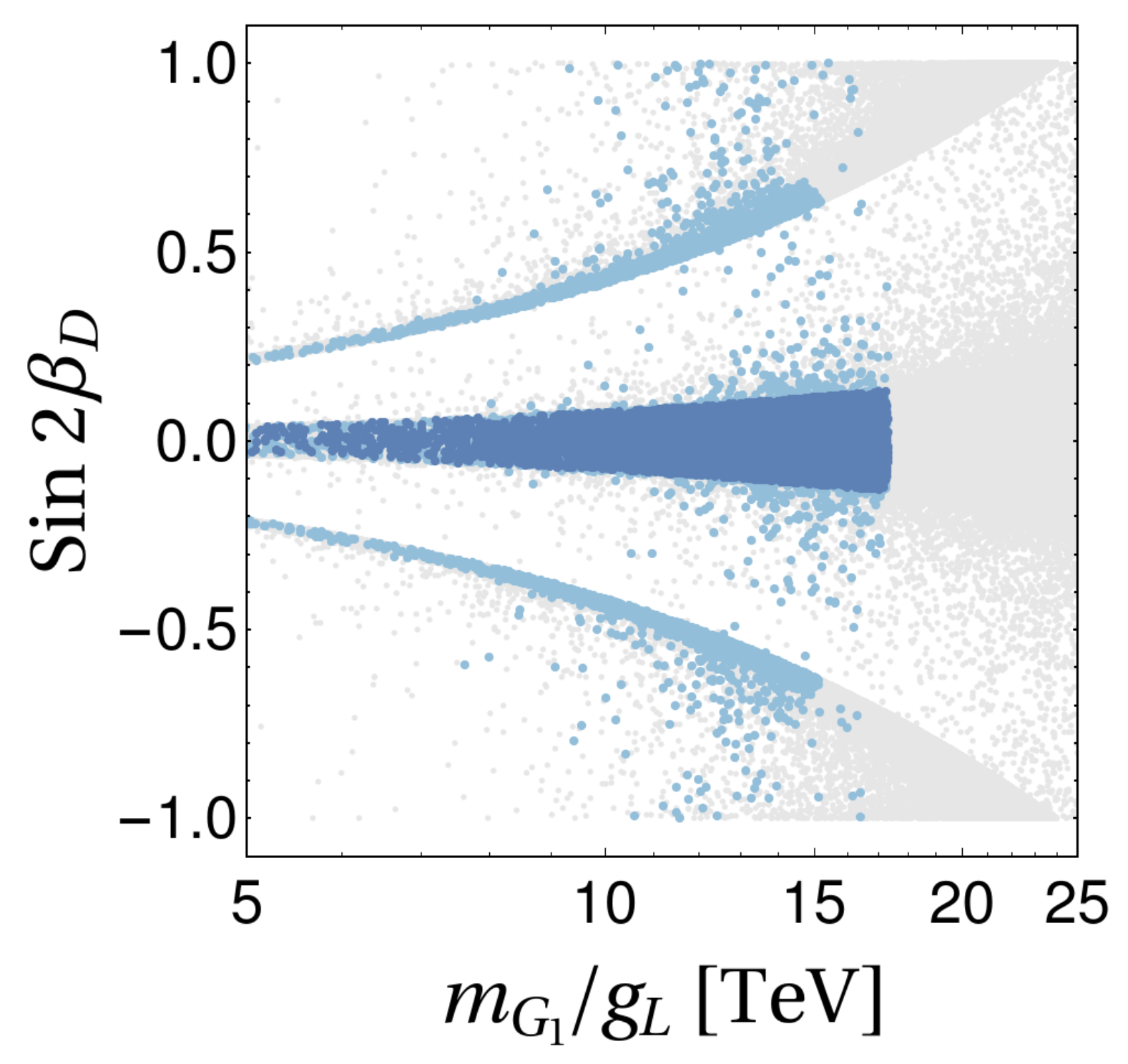}
\caption{\small \sl $\sin 2 \beta_{L,R}$ vs. $m_{G_{1}}/g_L$ as implied by Eq. (\ref{eq:parameters_scenario1}), see text for details. Color code as in Fig.~\ref{fig:scenario1}.
}
\label{fig:scenario1-angles}
\end{figure}

Within scenario 1, the contribution to $D^0 - \bar D^0$ is due to
\be
\delta \ms L_{\rm eff}^{UU} = - \sum_a \dfrac{g_L^2}{2 m_{G_a}^2} \big( \bar{U}' \gamma^\mu_L T^a U' \big)^2 \supset C_{(u c)^2} \left( \bar u \gamma^\mu_L c \right)^2~,
\ee
where, exploiting CKM hierarchies, we can write
\be
\label{eq:DMD_scenario1}
C_{(u c)^2} = - (V_{us} V_{cs}^*)^2 \dfrac{g_L^2}{8} \left( \frac{\sin^2(2 \beta_D)}{m_{G_{1}}^2} + \frac{\cos^2(2 \beta_D)}{m_{G_{3}}^2} \right) + O(\lambda^4)~.
\ee
with $\lambda$ the Wolfenstein parameter. While the approximate formula in Eq.~(\ref{eq:DMD_scenario1}) is very convenient to exhibit the mechanism at work, the numerical analysis includes the exact CKM dependence. In turn, the model's contribution to $\delta C_{9,10}^{\mu\mu}$ reads
\be
\label{eq:wc-mumu-scenario1}
\delta C_9^{\mu\mu} = - \delta C_{10}^{\mu\mu} =  \frac{g_L^2}{4} \frac{v^2 \pi}{\alpha_{\rm em} V_{tb} V_{ts}^*} \left[ \frac{\sin(2 \beta_D) \cos(2 \beta_L)}{m_{G_{3}}^2} - \frac{\sin(2 \beta_L) \cos(2 \beta_D)}{m_{G_{1}}^2} \right]~.
\ee
Clearly, with the advocated pattern of $G_{a}$ masses and rotation angles, the contribution in Eq.~(\ref{eq:DMD_scenario1}) will be parametrically suppressed by the decoupling of $G_{3}$ plus the smallness of $\beta_D$, and this very pattern ensures a sizeable contribution to $\delta C_9^{\mu\mu}$ from the second, negative term in Eq.~(\ref{eq:wc-mumu-scenario1}). These features are displayed more quantitatively in  Fig.~\ref{fig:scenario1}. In particular, the effective scale for the lighter among the $G_{a}$ bosons is shown versus $R_K$ in the left panel, where dark blue denotes points that fulfil all other constraints to be described later, and including $x_D$. The effective mass scale pointed to by the $x_D$ constraint, ruled by $m_{G_{3}}/g_L$,\footnote{It is clear that the requirement that $x_D$ saturate the experimental result entails a strong correlation between $m_{G_{1}}$ and $\beta_D$ in Eq.~(\ref{eq:DMD_scenario1}).} is displayed in the right panel of Fig.~\ref{fig:scenario1}.
In Fig. \ref{fig:scenario1-angles} we also display the hierarchy in the second of Eqs. (\ref{eq:parameters_scenario1}) versus the lightest $G_a$ mass. The left panel shows a quite strong correlation between the absolute scale of the effects and the actual size allowed to the large angle, namely $\beta_L$. This correlation arises from the measurement of $R_K^{(\ast)}$, in the limit of large $m_{G_3}$, as it can be seen from Eq. (\ref{eq:wc-mumu-scenario1}). Conversely, the right panel confirms that, once all constraints are taken into account, $\sin 2\beta_D$ should be small regardless of the value of $m_{G_1}$, as discussed below Eq.~(\ref{eq:parameters_scenario1}).

A few qualifications are now in order on the general setup of our numerical analysis. So far we focused on $D^0 - \bar D^0$ mixing because it turned out to be the most constraining observable within scenario 0. However, the pattern of parameters that we advocated in Eq.~(\ref{eq:parameters_scenario1}) may generate large effects in other observables, to be discussed in the next sections. With the exception of $\tau \to \mu \gamma$, that we comment upon in sec. \ref{sec:further_constraints}, all of our considered observables depend on the ratios $m_{G_{a}}/g_L$, rather than on masses and $g_L$ separately. Our main numerical scan then assumes the ranges to follow
\be
m_{G_{a}} / g_L \in [1, 500]~{\rm TeV} ~~~~~\&~~~~~ \beta_D, \beta_L \in [0,2\pi)~.  
\ee
It goes without saying that the upper limit of 500 TeV in general corresponds to $m_{G_{a}}$ values {\em well below} this mass scale, because $g_L$ is in general well below unity. As concerns the hierarchies in Eq.~(\ref{eq:parameters_scenario1}), we followed two alternative procedures: on the one side, we performed scans imposing such hierarchies from the outset; on the other side, we let the constraints choose them. We found no appreciable difference in the results obtained with these alternative procedures.

\subsection{\boldmath $b \to s \ell \ell'$ and leptonic LFV} \label{sec:LFV}

A large $\beta_L$, combined with a value for $m_{G_{1}}$ as low as required by $R_{K^{(*)}}$, may lead to troublesome effects in particular in $b \to s \tau^\pm \mu^\mp$ as well as in leptonic LFV decays such as $\tau \to 3 \mu$. Quite interestingly, the effects are indeed sizeable, but below existing limits. Besides, the small number of parameters involved establishes clear-cut correlations between LUV and LFV observables, as well as across different LFV observables. These correlations represent a prominent feature of our model, as can be qualitatively understood, again, from the basic formulae for the relevant Wilson coefficients. A first comment concerns $b \to s \tau \tau$. Since
\be
\delta C_{9,10}^{\tau \tau} = - \delta C_{9,10}^{\mu \mu}~,
\ee
the departure of $\mc B(B \to K \tau \tau)$ from its SM prediction can be written as a function of the departure of $R_K$ from unity. As a consequence, modifications of branching ratios for $B \to K \tau \tau$ as well as $B_s \to \tau \tau$ will be of the order of 20\% with respect to the respective SM expectations, which are sizeably below existing limits \cite{TheBaBar:2016xwe,Aaij:2017xqt}.

\begin{figure}[ht!]
\centering
	\includegraphics[width=0.55\linewidth]{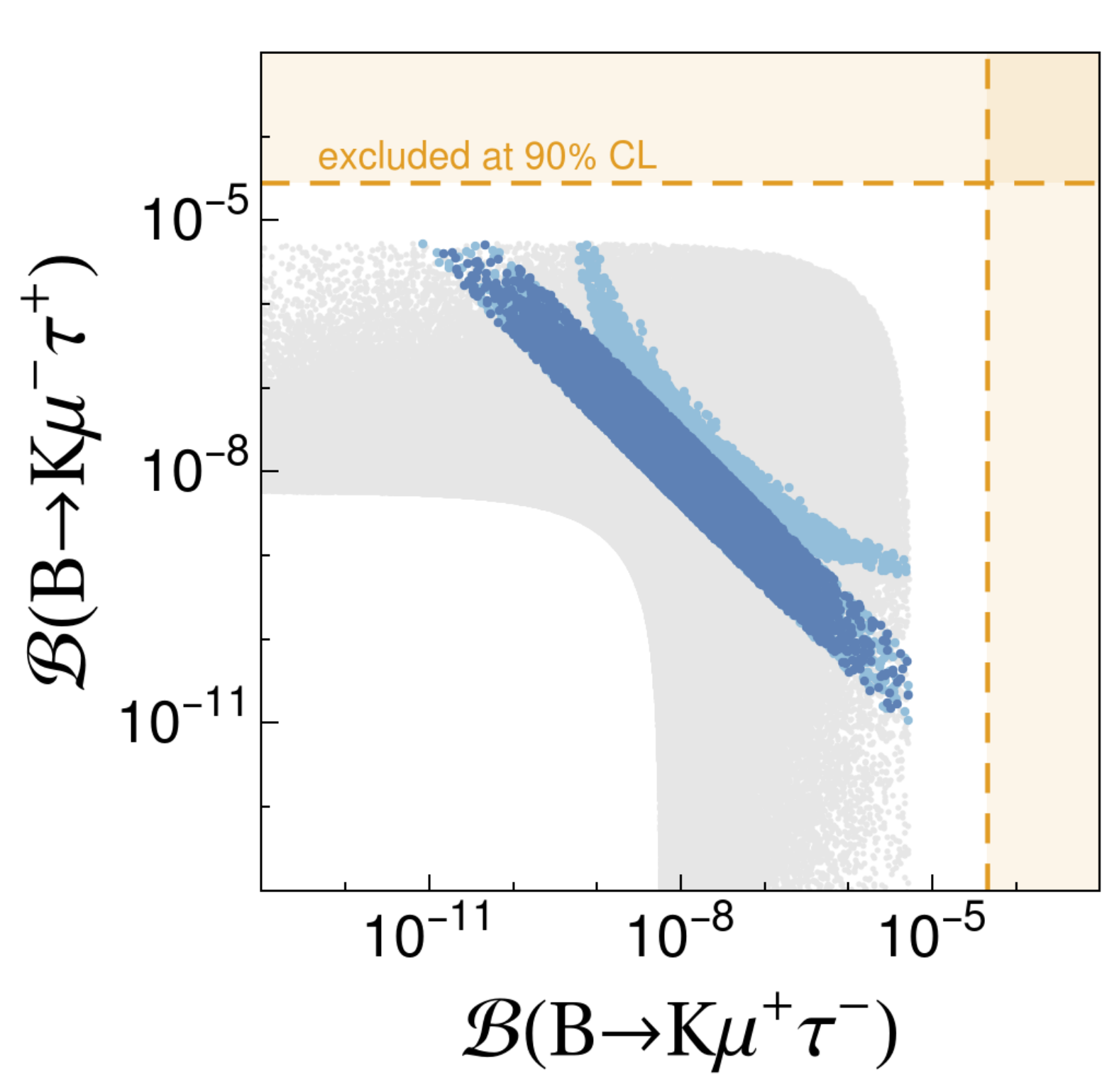}
\caption{\small \sl Correlation between the two $\mc B(B \to K \mu \tau)$ modes within scenario 1. Color code as in Fig.~\ref{fig:scenario1}. Dashed lines denote the existing bounds on the respective modes \cite{Lees:2012zz}, see text for details.}
\label{fig:BtoKmutau}
\end{figure}

We next turn to the predictions for $b \to s \tau \mu$ decays. The relevant Wilson coefficients read
\begin{align}
\label{eq:wc-mutau_scenario1}
\delta C_9^{\mu\tau} &= -\frac{g_L^2}{4} \frac{v^2 \pi}{\alpha_{\rm em} V_{tb} V_{ts}^*} 
\left( 
\frac{\cos2 \beta_D \cos2 \beta_L}{m_{G_{1}}^2}
+\frac{\sin2 \beta_D \sin2 \beta_L}{m_{G_{3}}^2} - \frac{1}{m_{G_{2}}^2}
\right)~, \\
\label{eq:wc-taumu_scenario1}
\delta C_9^{\tau\mu} &= -\frac{g_L^2}{4} \frac{v^2 \pi}{\alpha_{\rm em} V_{tb} V_{ts}^*} 
\left( 
\frac{\cos2 \beta_D \cos2 \beta_L}{m_{G_{1}}^2}
+\frac{\sin2 \beta_D \sin2 \beta_L}{m_{G_{3}}^2} + \frac{1}{m_{G_{2}}^2}
\right)~.
\end{align}
Keeping in mind the main assumptions defining our scenario 1, Eq.~(\ref{eq:parameters_scenario1}), it is clear that the dominant dependence is on $|\cos 2\beta_L \pm 1| / m_{G_{1,2}}^2$. Hence, a rather distinctive feature of this scenario is that $\mc B(B \to K \tau^+ \mu^-) \neq \mc B(B \to K \tau^- \mu^+)$, although either can be larger than the other, depending on the choice of the $\beta_L$ phase. The correlation between these two modes is displayed in Fig. ~\ref{fig:BtoKmutau}. The dominant parametric dependence highlighted above translates into an approximate reflection symmetry of the plot around the diagonal.

\begin{figure}[ht!]
\centering
	\includegraphics[width=0.50\linewidth]{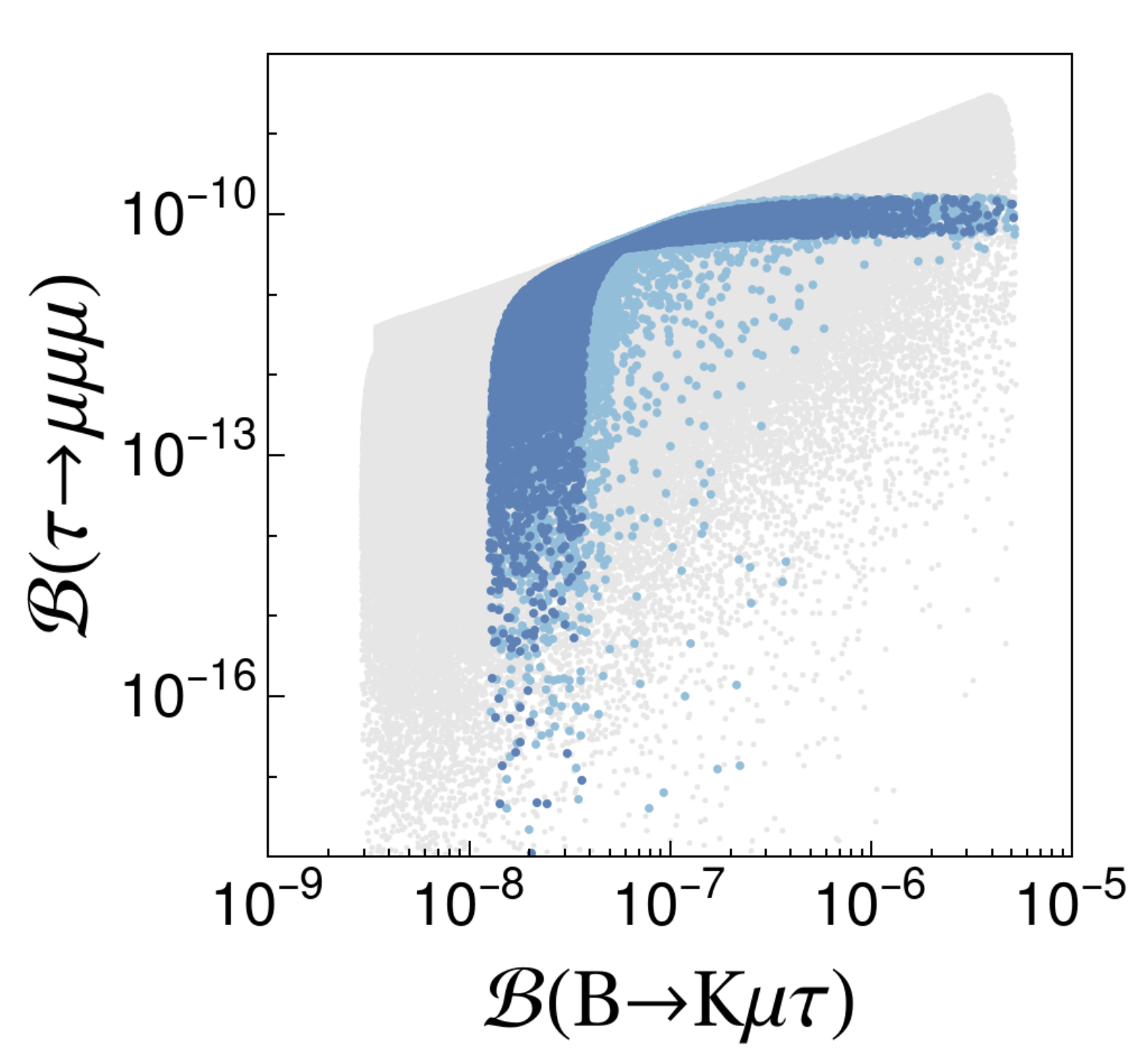}
	\hfill
	\includegraphics[width=0.48\linewidth]{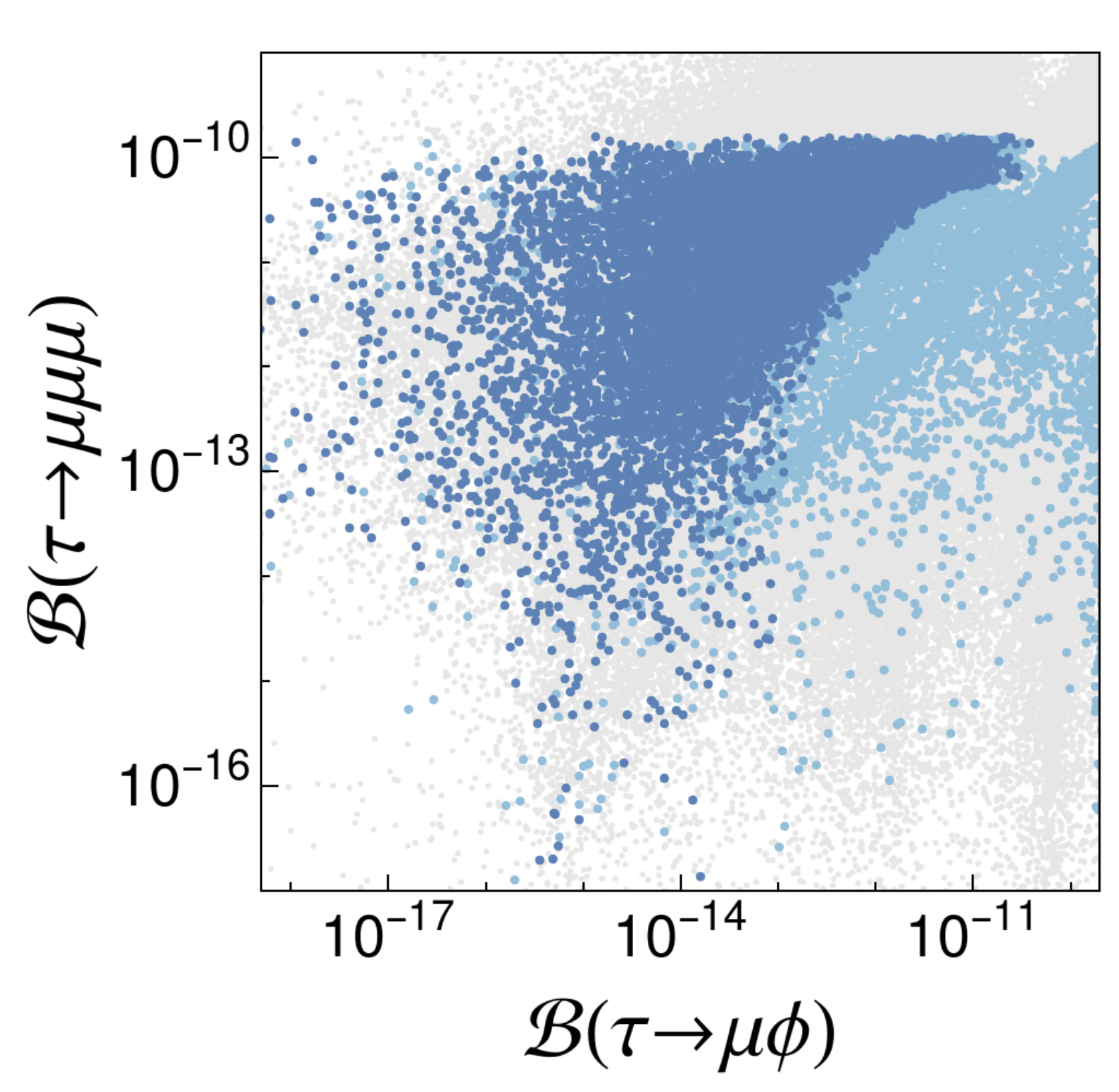}
\caption{\small \sl Left panel: correlation between $\mc B(B \to K \tau^\pm \mu^\mp)$ and $\tau \to 3 \mu$. Right panel: correlation between $\tau \to 3 \mu$ and $\tau \to \mu \phi$. Color code as in Fig.~\ref{fig:scenario1}.}
\label{fig:LFV_vs_LFV}
\end{figure}

Most importantly, our model predicts not only an upper bound, but also a lower bound on the LFV rates. 
We obtain (see also left panel of Fig.~\ref{fig:LFV_vs_LFV})
\be
1.3 \times 10^{-8}\lesssim \mc B(B\to K \mu^+ \tau^-)+\mc B(B\to K \mu^- \tau^+) \lesssim 5.2 \times 10^{-6}~.
\ee
Interestingly, the maximal rate predicted by our scenario lies just one order of magnitude below the existing limits obtained by BaBar \cite{Lees:2012zz}, $\mc B(B\to K \mu^+ \tau^-)_{\rm exp} < 4.5\times 10^{-5}$ and $\mc B(B\to K \mu^- \tau^+)_{\rm exp} < 2.8\times 10^{-5}$.\footnote{Also noteworthy, the lower bound is in good accord with the predictions obtained within approaches motivated by completely different considerations \cite{Guadagnoli:2015nra,Boucenna:2015raa}\cite{Becirevic:2016zri,Becirevic:2016oho,Bordone:2018nbg}.
} As discussed in Ref.~\cite{Becirevic:2016zri}, the predictions given above can be translated into corresponding fiducial ranges on the other exclusive decays generated by the $b \to s \mu\tau$ current
\be
\dfrac{\mc B(B_s \to \mu^\pm \tau^\mp)}{\mc B(B \to K \mu^\pm \tau^\mp)} \approx 0.9\,, \qquad\qquad \dfrac{\mc B(B \to K^\ast \mu^\pm \tau^\mp)}{\mc B(B \to K \mu^\pm \tau^\mp)}\approx 1.8\,,
\ee
where we take central values for the hadronic parameters. These relations show that $B \to (K) \ell \ell'$ decays with a final-state $\tau$ are expected to be related with each other by O(1) factors~\cite{Glashow:2014iga}, i.e.~that {\em experimentally reaching one of them will possibly lead to reaching them all.} Existing experimental limits for these channels are not very constraining \cite{Patrignani:2016xqp},\footnote{Strongest limits are on modes with final-state electrons and muons, e.g. Ref. \cite{Aaij:2017cza}.} and yet this is a very characteristic prediction of models that interpret $b \to s \ell \ell$ discrepancies as due to a $(V-A) \times (V-A)$ interaction coupled, above the EWSB scale, to third-generation-only SM fermions \cite{Glashow:2014iga}. 

As anticipated, the above LFV predictions are in turn correlated with purely leptonic LFV, in particular in the processes $\tau \to 3 \mu$ and $\tau \to \phi \mu$. From
\be
\delta \ms L_{\rm eff}^{LL} \supset + \dfrac{g_L^2}{8} \sin (4 \beta_L) \left( \frac{1}{m_{G_{3}}^2} - \frac{1}{m_{G_{1}}^2} \right) (\bar \tau \gamma^\mu_L \mu) (\bar \mu \gamma_{\mu L} \mu)~,
\ee
one gets \cite{Bertuzzo:2015ada}
\be
\mc B (\tau \to 3 \mu) = \frac{m_\tau^5}{3072 \pi^3 \Gamma_\tau} \frac{g_L^4}{64} \sin^2(4 \beta_L) \left( \frac{1}{m_{G_{3}}^2} - \frac{1}{m_{G_{1}}^2}\right)^2~.
\ee
Besides, from
\be
\delta \ms L_{\rm eff}^{LD} \supset C_{ss}^{\mu\tau} \left( \bar \mu \gamma^\mu_L \tau\right) \left( \bar s \gamma_{\mu \, L} s \right)
\ee
with
\be
C_{ss}^{\mu\tau} = - \dfrac{g_L^2}{4} \left[
- \frac{\cos (2 \beta_D) \sin (2 \beta_L)}{m_{G_{3}}^2} + \frac{\sin (2 \beta_D) \cos (2 \beta_L)}{m_{G_{1}}^2} 
\right]~,
\ee
one likewise arrives at
\be
\mc B (\tau \to \mu \phi) \simeq \left| C_{ss}^{\mu \tau} \right|^2 \frac{f_\phi^2 m_\phi^4}{64 \pi \, m_\tau \Gamma_\tau}\left( 1 - \frac{m_\phi^2}{m_\tau^2} \right) \left( -1 +\frac{m_\tau^2}{2 m_\phi^2} + \frac{m_\tau^4}{2 m_\phi^4}\right)~,
\ee
where $f_\phi=241(18)~{\rm MeV}$ is the $\phi$-meson decay constant~\cite{Donald:2013pea}, and we have neglected the $m_\mu$ mass dependence, which amounts to an approximation of few percent. 

Similarly to the transition $b\to s\mu^+\mu^-$, this observable is only modified for nonzero values of $\beta_L$. From the current experimental limit \cite{Miyazaki:2011xe,Schwanda:2014aca} (90\% CL), and keeping in mind Eq. (\ref{eq:parameters_scenario1}), we obtain
\be
\label{eq:bound-taumuphi}
\frac{m_{G_{1,2}}^2}{g_L^2 \sin 2 \beta_D \cos2 \beta_L} \ge (3.7~{\rm TeV})^2~,
\ee
which is once again consistent with the constraint derived in Eq.~\eqref{eq:RK-constraint} from $R_{K^{(\ast)}}$. From the requirement that the $R_{K^{(*)}}$ discrepancies be reproduced at 1$\sigma$, we obtain $\mc B(\tau \to \mu \phi)$ as large as $1 \times 10^{-10}$ and, in general, model points mostly populating the range between $10^{-14}$ and $10^{-10}$ (see right panel of Fig.~\ref{fig:LFV_vs_LFV}).
It is worth mentioning that the projected Belle-II sensitivity to this decay is around $10^{-9}$ \cite{BelleII-note}.

In the parameter space of Eq.~(\ref{eq:parameters_scenario1}), the above formulae translate into a triple correlation between $\mc B(B \to K \mu^\pm \tau^\mp)$, $\mc B(\tau \to 3 \mu)$ and $\mc B(\tau \to \mu \phi)$, illustrated in the two plots of Fig.~\ref{fig:LFV_vs_LFV}.

\subsection{$\Delta M_s / \Delta M_d$}

Mixings in the $B_{d,s}$ sector, in particular the ratio $\Delta M_s / \Delta M_d$, represent a strong constraint (see Ref. \cite{DiLuzio:2017fdq} for a recent discussion). Actually, it is mainly this constraint that selects the mass hierarchy in Eq.~(\ref{eq:parameters_scenario1}), the angles' hierarchy being instead mostly the result of $R_K$ and $x_D$. From the effective Hamiltonian
\be
\mc H_{\rm eff}^{\Delta M_q} = ( C_1^{\rm SM} + \delta C_1^q) (\bar q \gamma^\mu_L b)^2 + {\rm h.c.}
\ee
with $q$ either of $d$ or $s$, and using Ref. \cite{Buchalla:1995vs} to get $C_1^{\rm SM}$ in our normalization, we obtain
\be
\delta C_1^s = \frac{g_L^2}{8} \left( \frac{\cos^2 2 \beta_D}{m^2_{G_{1}}} + \frac{\sin^2 2 \beta_D}{m^2_{G_{3}}} - \frac{1}{m_{G_{2}}^2} \right)~,~~~~~~~ \delta C_1^d = 0~.
\ee
We can then write
\be
\frac{\Delta M_s}{\Delta M_d} = \left| \frac{V_{ts}}{V_{td}} \right|^2 \xi^2 \frac{m_{B_s}}{m_{B_d}} \left| 1 + \frac{\delta C_1^s}{C_1^{\rm SM}} \right|~.
\ee
Taking $\xi = 1.239(46)$ \cite{Aoki:2016frl} as well as $|V_{ts} / V_{td}| = 4.58(24)$ from a Unitarity-Triangle fit using only quantities not affected by new physics \cite{Derkach-private,UTfit},\footnote{A similar prediction may be obtained using the CKMfitter code \cite{CKMfitter}.} one would obtain the SM prediction $(\Delta M_s / \Delta M_d)_{\rm SM} = 33(4)$, perfectly consistent with the value $(\Delta M_s / \Delta M_d)_{\rm exp} = 35.06(14)$ obtained from the mass differences reported in Ref. \cite{Patrignani:2016xqp}. Our model's prediction for this ratio, normalized to the SM result, reads
\be
\label{eq:DMs/DMd}
\frac{\Delta M_s}{\Delta M_d} ~ \Big/ \left( \frac{\Delta M_s}{\Delta M_d} \right)_{\rm SM} \in [0.8,1]~.
\ee
In short, the model tends to predict a suppression of the order of 10-20\%. However, such shift comes with an error of comparable size, about 12\%, dominated by the CKM input, followed by the $\xi$ input. This error at present prevents this observable from providing a stringent test of our framework. Such test will however be possible with improvements on fits to the unitarity triangle using only observables realistically unaffected by new physics, such as $\gamma$ from $B \to D K^{(*)}$ (see in particular \cite{Brod:2013sga,Aaij:2018uns,Craik:2017dpc}). This highlights the well-known importance of improvements in such `standard-candle' measurements.

\subsection{$B\to K \nu \bar{\nu}$}

Within our model, an explanation of $R_{K^{(\ast)}}$ implies new contributions to the $B\to K^{(*)} {\nu}\bar{\nu}$ decays. The only part of the $b \to s \nu \bar \nu$ Hamiltonian relevant to our discussion is (we adhere to the notation in \cite{Buras:2014fpa})
\be
\label{eq:hamiltonian-bsvv}
\mc H_{\rm eff} = - \frac{4 G_F}{\sqrt{2}}V_{tb}V_{ts}^* ~ C_L^{ij} \mc O_L^{ij} + {\rm h.c.}
\ee
with
\be
\mc O_L^{ij} = \frac{e^2}{(4\pi)^2}(\bar s \gamma^\mu_L b)(\bar \nu_i \gamma_\mu (1 - \gamma_5) \nu_j)~.
\ee
We define the ratio $R_{\nu\nu}^{(*)} \equiv \mc B(B \to K^{(*)} \nu\nu)/\mc B(B\to K^{(*)} \nu\nu)^{\rm SM}$, where, as usual, a sum over the (undetected) neutrino species is understood. In our scenario this ratio is modified as follows
\begin{align}
\label{eq:Rnunu}
\begin{split}
R_{\nu\nu}^{(*)} = \dfrac{\sum_{ij} | \delta_{ij} C_L^{\rm SM}+ \delta C_L^{ij}|^2}{3|C_L^{\rm SM}|^2} = 1 + \dfrac{2 C_L^{\rm SM} \sum_i \delta C_L^{ii}+\sum_{ij} |\delta C_L^{ij}|^2}{3|C_L^{\rm SM}|^2} ~,
\end{split}
\end{align}
where $C_L^{\rm SM}=-6.38(6)$ is the SM Wilson coefficient as defined in Ref.~\cite{Buras:2014fpa}. The contributions to $R_{\nu\nu}^{(*)}$ induced by new physics are encoded in the Wilson coefficients $\delta C_L^{ij}$, which satisfy $\delta C_L^{ij}=2\, \delta C_9^{ij}$, cf.~Eq.~(\ref{eq:hamiltonian-bsll}). Note that $R_{\nu \nu} = R^{*}_{\nu \nu}$ in our framework because of the absence of contributions to the right-handed counterpart of the operator $\mc O_L^{ij}$ \cite{Hiller:2014ula}. By replacing Eqs.~(\ref{eq:wc-mumu})--(\ref{eq:wc-mutau}) in Eq.~(\ref{eq:Rnunu}), we obtain
\be
\label{eq:Rnunu-model}
R_{\nu\nu}^{(*)} = 1 + \dfrac{1}{3 |C_L^{{\rm SM}}|^2} \frac{v^4 \pi^2}{\alpha^2_{\rm em} |V_{tb} V_{ts}^*|^2}
\frac{g_L^4}{8} 
\left( 
\frac{\cos^2 2 \beta_D}{m_{G_{1}}^4} + \frac{\sin^2 2 \beta_D}{m_{G_{3}}^4} + \frac{1}{m_{G_{2}}^4}
\right)~.
\ee
Interestingly, the flavour-diagonal contributions satisfy $\delta C_L^{\mu\mu} = - \delta C_{L}^{\tau\tau}$ (whereas $C_L^{ee}$ is not modified), so that interference terms between the SM and NP {\em vanish}, and NP contributions only enter at second order in the small ratio $\delta C_L^{ij} / C^{\rm SM}_L$. Because of this feature, which is a consequence of the underlying $\suh$ symmetry, the strong experimental constraints $R_{\nu\nu}^\ast < 2.7$ \cite{Grygier:2017tzo,Buras:2014fpa} do not pose, within our model, a challenge in the description of $b \to s$ anomalies. More quantitatively, using Eq. (\ref{eq:parameters_scenario1}) the $R_{\nu\nu}^{\rm exp}$ constraint can be translated into the bound
\be
\label{eq:bound-bsnunu}
\frac{m^2_{G_{1,2}}}{g_L^2\, \sqrt{1 + \cos^2 2 \beta_D}} \ge (3.8~{\rm TeV})^2~,
\ee
much weaker than the constraint derived in Eq.~(\ref{eq:RK-constraint}). Note that the dependence on $\beta_L$ disappears because of the sum over all neutrino species.

\subsection{$\tau\to \mu\nu\bar{\nu}$}

In spite of the horizontal gauge bosons $G_{a}$ being electrically neutral, they can also contribute to processes that in the SM are generated by charged currents, for example $\ell \to \ell' \nu \bar \nu$ decays. More precisely, one can show that the following term appears in $\delta \ms L_{\rm eff}$
\begin{align}
\begin{split}
\delta \ms L_{{\rm eff}} &\supset -\dfrac{g_L^2}{4}\left( \frac{\sin^2 2 \beta_L}{m^2_{G_3}} + \frac{\cos^2 2 \beta_L}{m^2_{G_1}} +\frac{1}{m^2_{G_2}} \right)(\bar{\nu}_{\tau} \gamma^\mu_L \nu_\mu)(\bar{\mu} \gamma_{\mu \, L} \tau)\,.
\end{split}
\end{align}
This contribution entails the following modification of $\mc B(\tau \to \mu \nu \nu)$ with respect to its SM prediction 
\be
\dfrac{\mc B(\tau \to \mu \nu\nu)}{\mc B(\tau \to \mu \nu\nu)^{{\rm SM}}} = 1 + \frac{g_L^2 v^2}{4} \left( \frac{\cos^2 2 \beta_L}{m^2_{G_{1}}} + \frac{\sin^2 2 \beta_L}{m^2_{G_{3}}} + \frac{1}{m^2_{G_{2}}} \right) + ...~,
\ee
where, for simplicity, we only show the dominant term coming from the interference, whereas in the numerics we include also the subleading (new-physics)$^2$ contributions.
By using the experimental average $\mc B (\tau \to \mu \bar{\nu}_\mu \nu_\tau)_{{\rm exp}}=17.33(5)\%$~\cite{Patrignani:2016xqp} and the SM prediction $\mc B(\tau \to \mu \bar{\nu}_\mu \nu_\tau)^{{\rm SM}}=17.29(3)\%$~\cite{Becirevic:2016zri}, we obtain the following $2\sigma$ bound
\be
\label{eq:bound-taumununu}
\frac{m^2_{G_{1,2}}}{g_L^2 \, \sqrt{1 + \cos^2 2 \beta_L}} \ge (1.6~{\rm TeV})^2~,
\ee
which is weaker than the one derived in Eq.~\eqref{eq:bound-bsnunu} from the experimental limit on $\mc B(B\to K \nu\nu)$.

\subsection{$D^0\to \mu\mu$} \label{sec:D0mumu}

Similarly to $D^0 - \bar D^0$ mixing, the CKM matrix induces a nonzero contribution to other charm-physics observables, most notably $D^0 \to \mu \mu$. We will show however that the induced modifications are too small to be observed with the sensitivity of the current experiments. The piece of Eq.~(\ref{eq:dLeff_rotated}) describing four-fermion interactions of up-type quarks $U'$ and charged leptons $L'$, the primes denoting as usual the `gauge' basis, reads
\be
\delta \ms L_{\rm eff}^{UL} = - \sum_a \dfrac{g_L^2}{m_{G_a}^2} \big( \bar{U}' \gamma^\mu_L \, T^a U' \big)\big( \bar{L}' \gamma_{\mu \, L} T^a L' \big) \supset C^{\mu \mu}_{u c} \left( \bar u \gamma^\mu_L c \right) \left( \bar \mu \gamma_{\mu \, L} \mu \right)~.
\ee
After manipulations entirely analogous to those leading to Eq.~(\ref{eq:DMD_scenario1}), we obtain
\begin{align}
\label{eq:Dtomumu}
\begin{split}
C^{\mu \mu}_{u c} = - V_{cs}^* V_{us} \dfrac{g_L^2}{4} &
\left( \frac{\cos 2 \beta_D \cos 2 \beta_L}{m_{G_{3}}^2} +
\frac{\sin 2 \beta_D \sin 2 \beta_L}{m_{G_{1}}^2}
\right) + O(\lambda^3)~.
\end{split}
\end{align}
We note again that the expansion in $\lambda$ is only for illustrative purposes, and that in the numerics we use exact expressions. The corresponding branching ratio is then given by
\begin{align}
\label{eq:Dmumu}
\begin{split}
\mc B(D^0\to\mu^+\mu^-) &= \dfrac{f_D^2 m_\mu^2 m_D}{32 \pi \Gamma_D} \sqrt{1-\dfrac{4 m_\mu^2}{m_{D^0}^2}} \, \times \\
&\left[ \dfrac{g_L^2}{4} \left| V_{cs}^* V_{us} \right| \left( \frac{\cos 2 \beta_D \cos 2 \beta_L}{m^2_{G_{3}}} + \frac{\sin 2 \beta_D \sin 2 \beta_L}{m^2_{G_{1}}} \right) \right]^2 + O(\lambda^4)~.
\end{split}
\end{align}
where we have neglected the small SM contribution. This expression should be confronted with the current experimental limit $\mc B(D^0\to \mu^+\mu^-)_{{\rm exp}} < 6.2\times 10^{-9}$~\cite{Aaij:2013cza}. However, by inspection of Eq.~(\ref{eq:Dmumu}) one sees that the relevant masses being bounded are a combination of $m^2_{G_{1}} / \sin 2 \beta_D$ and $m^2_{G_{3}}$. Keeping in mind the parameter space in Eq.~(\ref{eq:parameters_scenario1}), one concludes that Eq.~(\ref{eq:Dmumu}) provides a bound on the heavier scale, not the lighter one, and is thus irrelevant.

\subsection{Further constraints} \label{sec:further_constraints}

Here we collect comments on further, potentially constraining, experimental information not discussed so far. A first comment deserves the decay $\tau \to \mu \gamma$, whose current experimental limit reads $\mc B(\tau \to \mu \gamma) \le 4.4 \times 10^{-8}$ \cite{Aubert:2009ag}. By construction, our model induces the required dipole interaction only at one loop. Therefore this decay is, at present, not very constraining within our model, as it suffers from a further loop and $\alpha_{\rm em}$ suppression with respect to the other $\tau$ LFV decays discussed before, whose predictions are summarised in the right panel of Fig.~\ref{fig:LFV_vs_LFV}.

Further consideration deserve possible bounds coming from direct searches. To our knowledge, the most relevant analysis for our case is Ref. \cite{Greljo:2017vvb}. In particular, our model induces a contribution to $pp(s \bar{s}) \to \mu \mu$, which can be tested at the LHC by looking at the tails of dilepton distributions. Assuming that such distortions be the result of contact interactions of the kind
\be
\ms L_{\rm eff} \supset \frac{C_{ij}^{D \mu}}{v^2} (\bar d^i \gamma^\mu_L d^j) (\bar \mu \gamma_{\mu L} \mu)~,
\ee
with an effective scale well above the typical momentum exchange in the process, Ref. \cite{Greljo:2017vvb} quotes a present-day limit on $|C_{ss}^{D\mu}|$ of around $1 \times 10^{-2}$. In our model, this coefficient is of order $g_L^2 \cdot v^2 / {\rm min}(m^2_{G_{a}})$, that we can bound with $1 \times (0.246 / 5)^2 \simeq 2 \times 10^{-3}$, see e.g. $x$-axis scale on the left panel of Fig.~\ref{fig:scenario1}. It is true that in some parts of our parameter space -- with very low ${\rm min}(m_{G_{a}})$ and sizeable $g_L$ -- there may be distortions with respect to the effective-theory description in Ref. \cite{Greljo:2017vvb}. While this aspect may warrant further investigation, we believe that the above argument provides a robust order-of-magnitude assessment of the constraint.

\section{Conclusions}

Semileptonic decays involving $b \to s$ and $b \to c$ quark currents display persistent deviations with respect to Standard-Model predictions, at present the only coherent array of departures from the Standard Model in collider data. The putative new dynamics must, directly or indirectly, involve the second and the third generation of quarks and leptons. Furthermore, it must produce sufficiently large effects in the product of a quark times a charged-lepton bilinear, $J_q \times J_\ell$, and sufficiently small effects in flavour-changing $J_q \times J_q$ and $J_\ell \times J_\ell$ amplitudes. Especially this second requirement has greatly oriented the model-building literature towards leptoquark models, that avoid the problem by construction, although they are not free of other shortcomings. In this paper we take a different approach. The two aforementioned requirements invite consideration of a ÔhorizontalÕ group, $SU(2)$ being the smallest one that may be at play. We accordingly invoke the possibility of a gauged such symmetry, $\suh$, with all the left-handed $2^{\rm nd}$- and $3^{\rm rd}$-generation fermions universally charged under the corresponding group -- in the `gauge' basis.

After integrating out the heavy $\suh$ bosons, one generates all sorts of $J_{q,\ell} \times J_{q,\ell}$ amplitudes. However, assuming degenerate masses for the horizontal bosons, and in the absence of mixing between the two heavier generations and the lighter one, the assumed symmetry would make $J_q \times J_q$ and $J_\ell \times J_\ell$ amplitudes exactly flavour-diagonal, in the fermion mass eigenstate basis. This property prevents dangerous tree-level contributions to processes such as meson mixings and purely leptonic flavour-violating transitions. In reality, such contributions are not exactly zero because of CKM-induced mixing across all the generations. The most constraining of these effects turns out to be the mass difference in the $D^0 - \bar D^0$ system, $\Delta M_D$.

However, and quite remarkably in our view, one can accomplish a successful description of $b \to s$ deviations as well as of all constraints, by advocating a splitting of the horizontal-boson masses -- per se a plausible possibility -- in particular a configuration with two mass-degenerate gauge bosons hierarchically lighter than the third one.

Our scenario has, by construction, distinctive signatures in $b \to s \ell \ell'$ decays. In particular, $B \to K \tau^\pm \mu^\mp$ is predicted in the range
\be
1.3 \times 10^{-8}\lesssim \mc B(B\to K \mu^+ \tau^-)+\mc B(B\to K \mu^- \tau^+) \lesssim 5.2 \times 10^{-6}~,
\ee
with the $\tau^+ \mu^-$ and $\tau^- \mu^+$ modes in general differing by a sizeable amount that could have either sign. Besides, the small number of parameters involved establishes clear-cut correlations between semi-leptonic LFV decays of $B$ mesons and LFV decays involving only leptons, in particular a triple correlation between $B(B \to K \mu^\pm \tau^\mp)$, $B(\tau \to 3 \mu)$ and $B(\tau \to \mu \phi)$.

The framework we advocate opens several follow-up directions. First, and needless to say, in order to be fully calculable beyond tree level, the model still requires specification of the scalar sector that accomplishes the spontaneous breaking of the $\suh$ symmetry. Addressing this question in full introduces a degree of model dependence, and probably requires more data. We believe that, for the sake of the present paper, the existence of such a scalar sector is just sufficient.

A central issue is whether an appropriate variation of the mechanism may also explain charged-current discrepancies in $b \to c \ell \nu$. We do not see how such an extension could avoid introducing relations between up-type and down-type fermion chiral rotations. Our main perplexity is in the fact that a quantitative change in the $R_{D^{(*)}}$ anomalies may change such relations {\em qualitatively}. Hence, as also commented upon in the Introduction, we refrained from attempting a unified description of all anomalies in this work.

Finally, another interesting question is whether a suitable extension of our framework may include a candidate for thermal Dark Matter. To this end, and although not required to cure anomalies in our framework, one may introduce additional matter, made stable by a $Z_2$ remnant of the spontaneously broken gauge symmetry, along the lines of, e.g., Ref. \cite{Cline:2017lvv}.

\section*{Acknowledgments}

We would like to thank Denis Derkach for important feedback on CKM-matrix input, as well as Luca Di Luzio and Marco Nardecchia for valuable comments. We also acknowledge useful exchanges with Damir Becirevic, Svjetlana Fajfer, Rabindra N. Mohapatra and Maurizio Pierini. The work of DG is partially supported by the CNRS grant PICS07229. This project has received support from the European Union's Horizon 2020 research and innovation programme under the Marie Sklodowska-Curie grant agreement N$^\circ$~674896.

\bibliographystyle{JHEP}
\bibliography{horizontal-note}

\end{document}